\documentclass[usegraphicx,usenatbib,useAMS]{mn2e}

\usepackage{times} 
\usepackage{aas_macros}     
\usepackage{amssymb}        

\newcommand{\gsim}{\ga}
\newcommand{\lsim}{\la}

\newcommand{\hMsol}{h^{-1}M_\odot}

\newcommand{\mum}{\mu{\rm m}}

\newcommand{\Mpc}{{\rm Mpc}}

\newcommand{\kms}{{\rm km\,s^{-1}}}

\newcommand{\Gyr}{{\rm Gyr}}

\newcommand{\mJy}{{\rm mJy}}

\newcommand{\galform}{{\tt GALFORM}} 
\newcommand{\grasil}{{\tt GRASIL}}

\title[The role of SMGs in hierarchical galaxy formation]{The role of submillimetre galaxies in hierarchical galaxy formation.}

\author[J. E. Gonz\'{a}lez et al.]{Juan
E. Gonz\'{a}lez$^{1}$\thanks{E-mail: jegonzal@eso.org}
C. G. Lacey$^{1}$, C. M. Baugh$^{1}$, C. S. Frenk$^{1}$\\
$^{1}$Institute for Computational Cosmology,
Department of Physics, Durham University, South Road, Durham DH1 3LE,
UK}

\begin{document}

\date{}

\pagerange{\pageref{firstpage}--\pageref{lastpage}} \pubyear{2009}

\maketitle
 
\label{firstpage}

\begin{abstract}
We study the role of submillimetre galaxies (SMGs) in the galaxy
formation process in the $\Lambda$~Cold Dark Matter cosmology. We use
the \citeauthor{Baugh05} semi-analytical model, which matches the
observed SMG number counts and redshift distribution by assuming a
top-heavy initial mass function (IMF) in bursts triggered by galaxy
mergers. We build galaxy merger trees and follow the evolution and
properties of SMGs and their descendants. Our primary sample of model
SMGs consists of galaxies which had 850$\mum$ fluxes brighter than
5$\mJy$ at some redshift $z>1$.  Our model predicts that the
present-day descendants of such SMGs cover a wide range of stellar
masses $\sim 10^{10} - 10^{12} \hMsol$, with a median $\sim
10^{11}h^{-1} M_{\odot}$, and that more than 70\% of these descendants
are bulge-dominated. More than 50\% of present day galaxies with
stellar masses larger than $7\times 10^{11}h^{-1}M_{\odot}$ are
predicted to be descendants of such SMGs. We find that although SMGs
make an important contribution to the total star formation rate at
$z\sim 2$, the final stellar mass produced in the submillimetre phase
contributes only 0.2\% of the total present-day stellar mass, and 2\%
of the stellar mass of SMG descendants, in stark contrast to the
popular picture in which the SMG phase marks the production of the
bulk of the mass of present day massive ellipticals.
\end{abstract}


\section{INTRODUCTION}

Our understanding of the star formation history of the Universe has
changed dramatically as new populations of star-forming galaxies have
been discovered at high redshifts. A key advance was the discovery of
the cosmic far-infrared background by COBE \citep{Puget96}, with an
energy density similar to that in the UV/optical background, which
implies that the bulk of star formation over the history of the
Universe has been obscured by dust, with most of the radiation from
young stars being reprocessed by dust grains to far-infrared and
sub-mm wavelengths. This was followed by the discovery of the
submillimetre galaxies (SMGs) using the Submillimetre Common User
Bolometer Array (SCUBA) instrument on the James Clerk Maxwell
Telescope (JCMT) \citep{Smail97,Hughes98}, which are generally
interpreted as dust-enshrouded galaxies undergoing a
starburst. Follow-up studies have concentrated on SMGs with 850$\mum$
fluxes brighter than about 5$\mJy$. Spectroscopy has revealed that
the bulk of such SMGs lie at redshifts $z \sim 1-3$, with a median around
$z\sim 2$, when the Universe had only 20\% of its current age
\citep{Chapman05}. Explaining the abundance and redshifts of SMGs has
posed a challenge to theoretical models of galaxy formation ever since
their discovery.

In the standard picture, due to the high star formation rates inferred
in these galaxies (of $\sim 100-1000 M_{\odot}yr^{-1}$ for a standard
initial mass function, e.g. \citealt {Hughes98,Chapman04}), the SCUBA
phase is expected to play an important role in the build-up of the
stars in massive galaxies. Central to this topic are the questions: do
SMGs represent the assembly of massive present day ellipticals? What
is the typical duration of the SCUBA phase? What are the typical
stellar and host halo masses of their descendants? Do the descendants
all have high masses, and are they preferentially spheroid-dominated?
And finally, what is the overall contribution of the SCUBA phase to
the stellar mass of present-day galaxies?

In this paper we address these questions using the predictions of a
semi-analytical model for galaxy formation and evolution. A powerful
feature of semi-analytical models is the ability to connect high
redshift galaxies to their present-day descendants
\citep{Baugh06}. Since it is believed that the dust plays a central
role in producing the submillimetre radiation observed from SMGs, the
model must take into account the energy reprocessed by dust when the
spectral energy distribution (SED) is calculated.

The analysis in this paper is based on the \citet{Baugh05} version of
the {\tt GALFORM} semi-analytical model. \citeauthor{Baugh05} found it
necessary to assume a top-heavy initial mass function (IMF) for
starbursts in order to reproduce the observed number counts and
redshift distribution of the submillimetre galaxies, while also being
consistent with other observational constraints, including the
luminosity function of Lyman-break galaxies at $z\sim 3$, the
present-day optical and near- and far-infrared luminosity functions,
and the gas fractions and metallicities of present-day
galaxies. {In the \citeauthor{Baugh05} model, the SMGs seen in
current surveys are mostly starbursts triggered by major and minor
mergers between gas-rich disk galaxies.}  This model was subsequently
shown to predict evolution of galaxies in the IR in good agreement
with {\em Spitzer} observations \citep{Lacey08}, and has also been
used to make predictions for {\em Herschel} \citep{Lacey10}. The model
also reproduces the local size-luminosity relation for late-type
galaxies, and the fractions of early and late-type galaxies
\citep{Gonzalez09}. \citet{Gonzalez10} perform a similar analysis to
the one presented in this paper, but for Lyman-break galaxies, rather
than SMGs.  Lyman-break galaxies are identified using photometric
selection to isolate the Lyman-break feature present in the rest-frame
ultraviolet spectral energy distribution of star-forming galaxies (and
which is accentuated due to absorption by gas along the line of sight
to the galaxy).  \citet{Lacey10a} present a detailed study of the
nature of Lyman-break galaxies in hierarchical galaxy formation
models.

The semi-analytical galaxy formation approach has been used by other
authors to model submillimetre galaxies. \citet{Kaviani03} were able
to reproduce the number counts of SCUBA sources, but only by treating
the dust temperature as a free parameter, for which they needed to
choose quite a low value (20-25 K), in apparent conflict with
subsequent observations of sub-mm SED shapes \citep{Coppin08}.
\citet{Fontanot07} reproduced the number counts of galaxies in the
$850\mu m$ band using a self-consistent calculation of the dust
temperature and a standard Salpeter initial mass function for all
types of star formation, but their model does not match the observed
redshift distribution of SMGs. \citet{Swinbank08} tested the
\citet{Baugh05} model by comparing in detail the SEDs and stellar,
dynamical, gas and halo masses of submillimetre galaxies against
observational data, finding broad agreement.  {Recently
\citet{Dave10} have proposed an alternative model of SMGs based on
hydrodynamical simulations, in which SMGs are massive galaxies
undergoing quiescent star formation with a normal IMF. However, they
do not calculate the sub-mm luminosities of their simulated galaxies,
and so do not show that their model can reproduce the observed number
counts and redshift distribution of SMGs.} Prior to the present paper,
no attempt has been made to relate the SMGs predicted in models with
their present-day descendants.

This paper is laid out as follows. In Section~2, we outline the galaxy
formation model {\tt GALFORM} used to predict the properties of
SMGs. In Section~3, we present examples from the model of galaxy
evolution histories which produce SMGs. In Section~4, we study
properties predicted for SMGs at different redshifts, such as stellar
and host halo masses and morphologies, and also the duration of the
SMG phase. In Section~5, we show the model predictions for the
present-day descendants of SMGs, including their stellar and host halo
masses and morphologies.  In Section~6, we compute what fraction of
present-day galaxies had SMG progenitors, and in Section~7 we
calculate the contribution of SMGs to the present-day stellar mass
density. Finally we present our conclusions in Section~8.

\section{Galaxy Formation Model}
\subsection{Basic components}

We use the Durham semi-analytical galaxy formation model, {\tt
GALFORM}, which is described in detail by \citet{Cole00} and
\citet{Benson03}.  The model is set in the Cold Dark Matter (CDM)
cosmology, with density parameter, $\Omega_{0}=0.3$, a cosmological
constant, $\Omega_{\lambda}=0.7$, Hubble constant $H_0 = 70
\kms\Mpc^{-1}$, baryon density, $\Omega_{b}=0.04$ and a power spectrum
normalization given by $\sigma_{8}=0.93$.  Galaxies in the model are
assumed to form inside dark matter haloes, and their subsequent
evolution is controlled by the merging histories of the host dark
matter haloes.  The physical processes modelled include: i) the
hierarchical assembly of dark matter haloes; ii) the shock heating and
virialization of gas inside the gravitational potential wells of dark
matter haloes; iii) the radiative cooling and collapse of the gas to
form a galactic disk; iv) star formation in the cool gas; v) the
heating and expulsion of cold gas through feedback processes such as
stellar winds and supernovae; vi) chemical evolution of the gas and
stars; vii) mergers between galaxies within a common dark halo as a
result of dynamical friction; viii) the evolution of the stellar
populations using population synthesis models; ix) the extinction and
reprocessing of starlight by dust.

The merger histories of dark matter haloes are calculated using a
Monte Carlo technique, following the formalism of extended Press \&
Schechter theory \citep{PressSchechter74,LaceyCole93,Cole00}. The
formation and evolution of a representative sample of dark matter
haloes is modelled.

We use the same galaxy formation parameters adopted by
\citet{Baugh05}, but we use a modified version of the Monte Carlo
merger trees \citep{Parkinson08}, which better reproduces the
properties of dark matter halo merger trees extracted from the
Millennium Simulation \citep{Springel05}. This is necessary because we
wish to connect high redshift galaxies with their present-day
descendants, and so we need a prescription for building merger
histories of dark matter haloes which is accurate over a long interval
in time. The \citeauthor{Parkinson08} algorithm for building halo
merger trees is a modified version of that introduced by \cite{Cole00}
and used in \cite{Baugh05}, and has been tuned to match the merger
histories extracted from the Millennium Simulation.  This modification
does not significantly affect the published predictions of the
\citeauthor{Baugh05} model, since those did not depend on the accuracy
of the trees over long time intervals. Instead, \citeauthor{Baugh05}
laid down grids of halos at a range of redshifts to compute the number
counts of SMGs and the cosmic star formation history. We have checked
that using the \citeauthor{Parkinson08} tree algorithm does not alter
the predictions presented in \citeauthor{Baugh05}

The \citet{Baugh05} model uses feedback from a superwind to suppress
the formation of massive galaxies and uses a top-heavy stellar IMF in
starbursts triggered by galaxy mergers to reproduce the observed
numbers of submillimetre and Lyman-break galaxies. An alternative
model was presented by \citet{Bower06}, which adopts a standard IMF
throughout, but uses AGN feedback to reduce the number of massive
galaxies. However, the \citeauthor{Bower06} model underpredicts the
number counts of submillimetre by more than an order of magnitude, so
we do not consider it further here. For a complete list of the
differences between these two models, see \citet{Gonzalez09}.

\subsection{Mergers and star formation bursts}

When dark matter haloes merge, the galaxy in the most massive halo is
assumed to become the central galaxy in the new halo while the other
galaxies become satellites (note that the central galaxy identified in
this way is not necessarily the most massive galaxy in the new halo).
The model assumes that both central and satellite galaxies can form
stars from their cold gas reservoirs, but only central galaxies can
accrete more cold gas by cooling from the surrounding halo. {
(\citet{Font08} considered an alternative model in which the hot gas
halos of satellite galaxies are only partly stripped, but this has
little effect on SMGs in our model, which are typically central
galaxies.)}


Mergers in the model are assumed to happen only between satellite and
central galaxies within the same halo.
Bursts of star formation in the model are triggered by galaxy mergers.
Two types of mergers are defined, major mergers and minor mergers,
according to whether or not the ratio of the mass of the smaller
galaxy to the larger galaxy $M_2/M_1$ exceeds the threshold $f_{\rm
ellip}$. Bursts are assumed to be triggered in all major mergers, but
also in minor mergers that satisfy $M_2/M_1 > f_{\rm burst}$ and where
the ratio of gas to stellar plus gas mass of the larger galaxy exceeds
$f_{\rm gas,crit}$.  The values adopted in the \citeauthor{Baugh05}
model are $f_{\rm ellip}=0.3$, $f_{\rm burst}=0.05$ and $f_{\rm
gas,crit}=0.75$.  Mergers can change the morphologies of galaxies.  In
a major merger, the stellar disks of the merging galaxies are assumed
to be transformed into a new stellar spheroid (resulting in a pure
bulge galaxy) and any gas present in the disks is converted into stars
in the burst, or ejected by supernova feedback associated with the
burst. In minor mergers, the stellar disk of the larger galaxy is
preserved and the stars from the smaller galaxy are added to its
bulge, so the merger remnant is a disk+bulge galaxy. If the condition
for a burst in this minor merger is satisfied, then the gas of the
smaller galaxy is converted into stars and added to the bulge, while
if the condition is not satisfied, any gas is added to the main gas
disk.

{As shown in \citet{Baugh05}, our model predicts that most SMGs
seen in current sub-mm surveys are starbursts triggered by major and
minor mergers involving gas-rich disk galaxies. \citet[][and
references therein]{Dekel09} have proposed a picture in which the most
actively star forming galaxies at high redshifts are fed by cold
streams of gas falling into galaxy halos. Although this picture
appears very different from our model, the differences are actually
less dramatic than they appear at first sight. Firstly, the \galform\
semi-analytical model has always included a ``cold'' mode of
accretion, in which gas falls to the halo centre at the free-fall
rate, for halos in which the cooling time is less than the free-fall
time for gas shocked at the virial radius \citep{Cole00}. This
``cold'' mode dominates at lower halo masses and higher
redshifts. \citet{Benson10} have recently shown that using the
\citet{Birnboim03} criterion to distinguish ``cold'' and ``hot''
accretion modes instead of the standard \galform\ criterion slightly
shifts the halo mass at which the transition from ``cold'' to ``hot''
accretion occurs, but leaves gas accretion rates and SFRs in the model
almost unchanged. They also find that the importance of the ``cold''
mode is greatly reduced when realistic feedback is included. Secondly,
in the \citeauthor{Dekel09} picture, the bright SMGs are actually
triggered by the infall of large clumps of cold gas rather than smooth
streams -- effectively gas-rich galaxy mergers. }

\subsection{Top-heavy IMF in bursts}
\label{ssec:IMF}

As mentioned in the introduction, an important feature of the
\citeauthor{Baugh05} model is that stars are assumed to form with
different initial mass functions (IMFs) in different
environments. This differentiation depends on whether they form
quiescently in disks or in bursts following a galaxy merger. In the
case of disks, the model adopts a standard solar neighbourhood IMF
proposed by \citet{Kennicutt83}, for which the number of stars
produced varies with mass as ${\rm d}N/{\rm d}\ln m \propto m^{-x}$,
with $x=0.4$ for $m < 1 M_{\odot}$ and $x=1.5$ for $m > 1
M_{\odot}$. In the case of bursts, a top-heavy IMF is used, where
$x=0$. In either case, the IMF covers a mass range $0.15 < m < 120
M_{\odot}$. This top-heavy IMF in bursts has two effects in the model:
i) it increases the total UV radiation from the young stars formed and
ii) it increases the metal production and hence also the amount of
dust.

When stars die, they return gas and metals to the interstellar medium
(ISM).  We treat this process using the instantaneous recycling
approximation, in which the metal ejection is quantified by the yield,
$p$, (which is the mass of metals returned to the ISM per unit mass of
stars formed) and the gas return is specified by the recycled
fraction, $R$, (defined as the mass of gas returned to the ISM per
unit mass of stars formed). We have calculated the values of $p$ and
$R$ for our two IMFs based on the results of stellar evolution
calculations \citep{Cole00,Baugh05}, and find values $p_{\rm quiescent}=0.023$,
$R_{\rm quiescent}=0.41$ for quiescent star formation  and
$p_{\rm burst}=0.15$, $R_{\rm burst}= 0.91$ for bursts.

There is independent support for the idea of variations in the IMF
from a variety of observations, as discussed in \citet{Lacey08}. We
mention a few of these here.  \citet{Fardal07} found a discrepancy
between the observed ratio of the total extragalactic background
radiation to the observed stellar density today and the value
predicted using the observationally inferred star formation history,
if a standard IMF is assumed. They find that a more top-heavy IMF is
needed (increasing the proportion of stars formed with $1.5 M_{\odot}
< m < 4 M_{\odot})$ in order to reconcile these different
observations. They argue that alternative solutions in which
non-stellar sources make large contributions to the background light
appear unlikely to resolve the discrepancy.  Another suggestion of
variations in the IMF comes from the work of \citet{vanDokkum08}, who
compares the luminosity evolution of massive cluster galaxies (at
$0.02 \leq z \leq 0.83$) with their colour evolution. He finds that
the evolution of the rest-frame $U-V$ colour is not consistent with
the previously determined evolution of the rest-frame $M/ L_{B}$ ratio
for a standard IMF, but shows that using a different IMF with a
flatter slope compared with a standard IMF can help to solve the
discrepancy.

\begin{figure*}
\includegraphics[width=13.5cm]{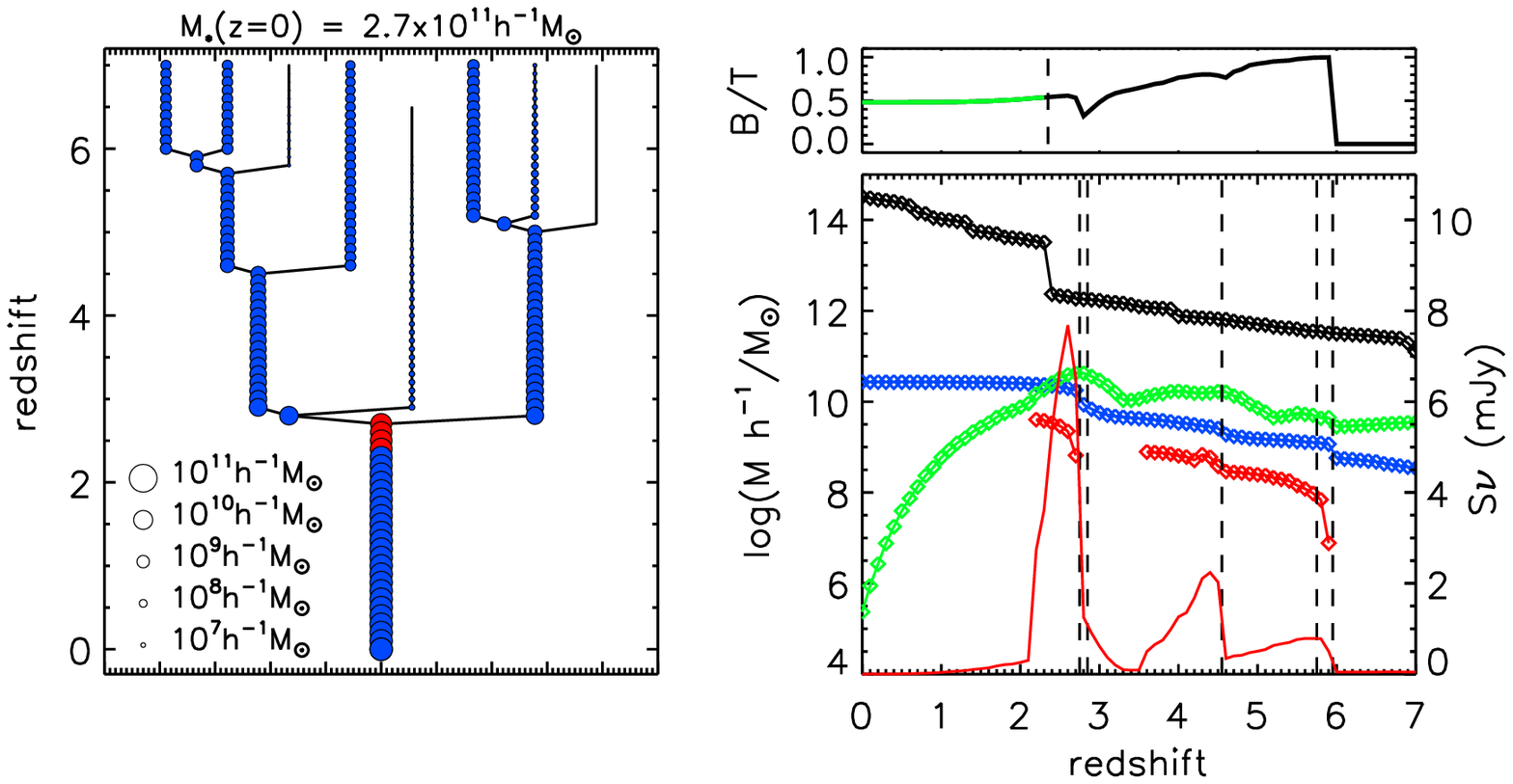}
\includegraphics[width=13.5cm]{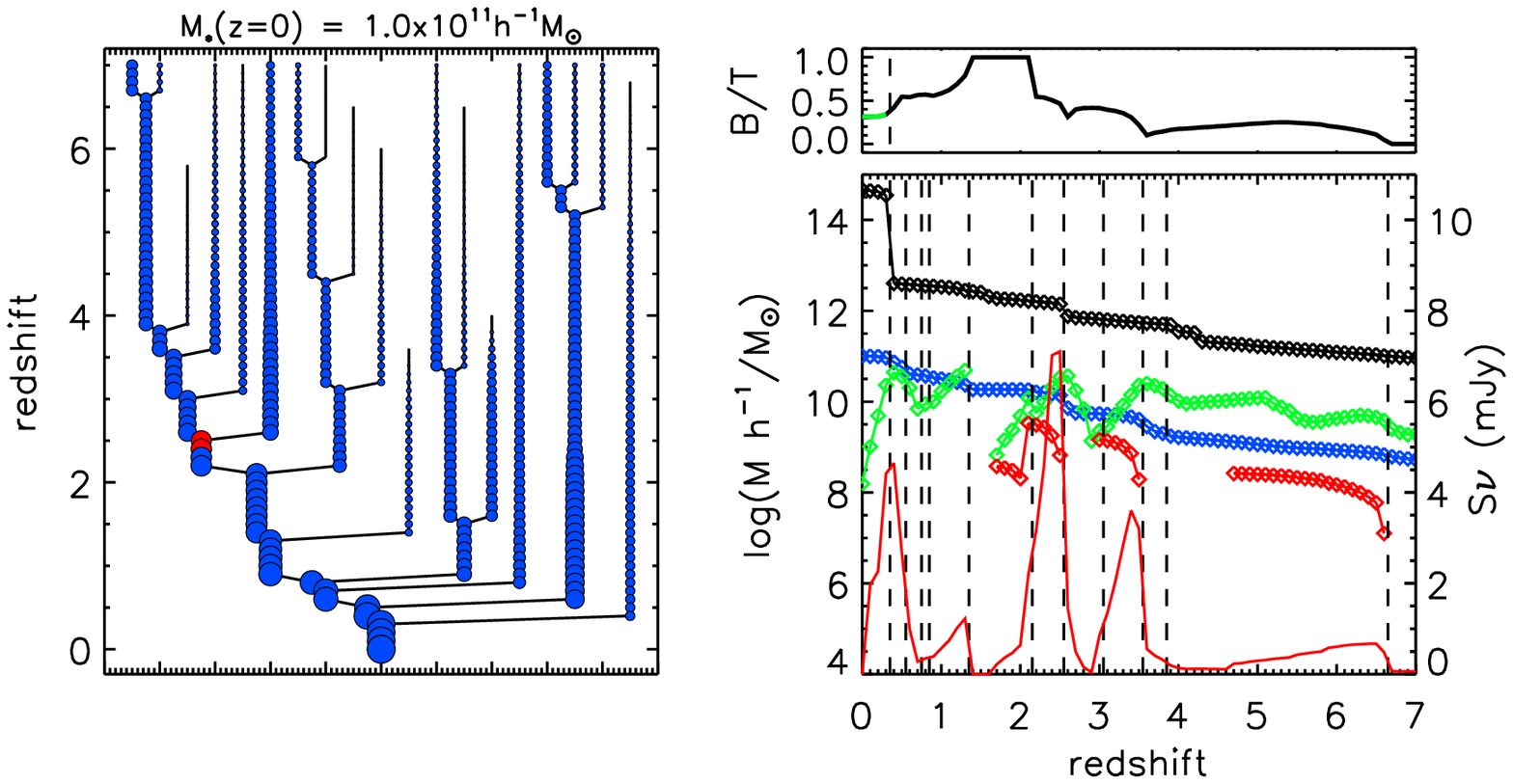}
\includegraphics[width=13.5cm]{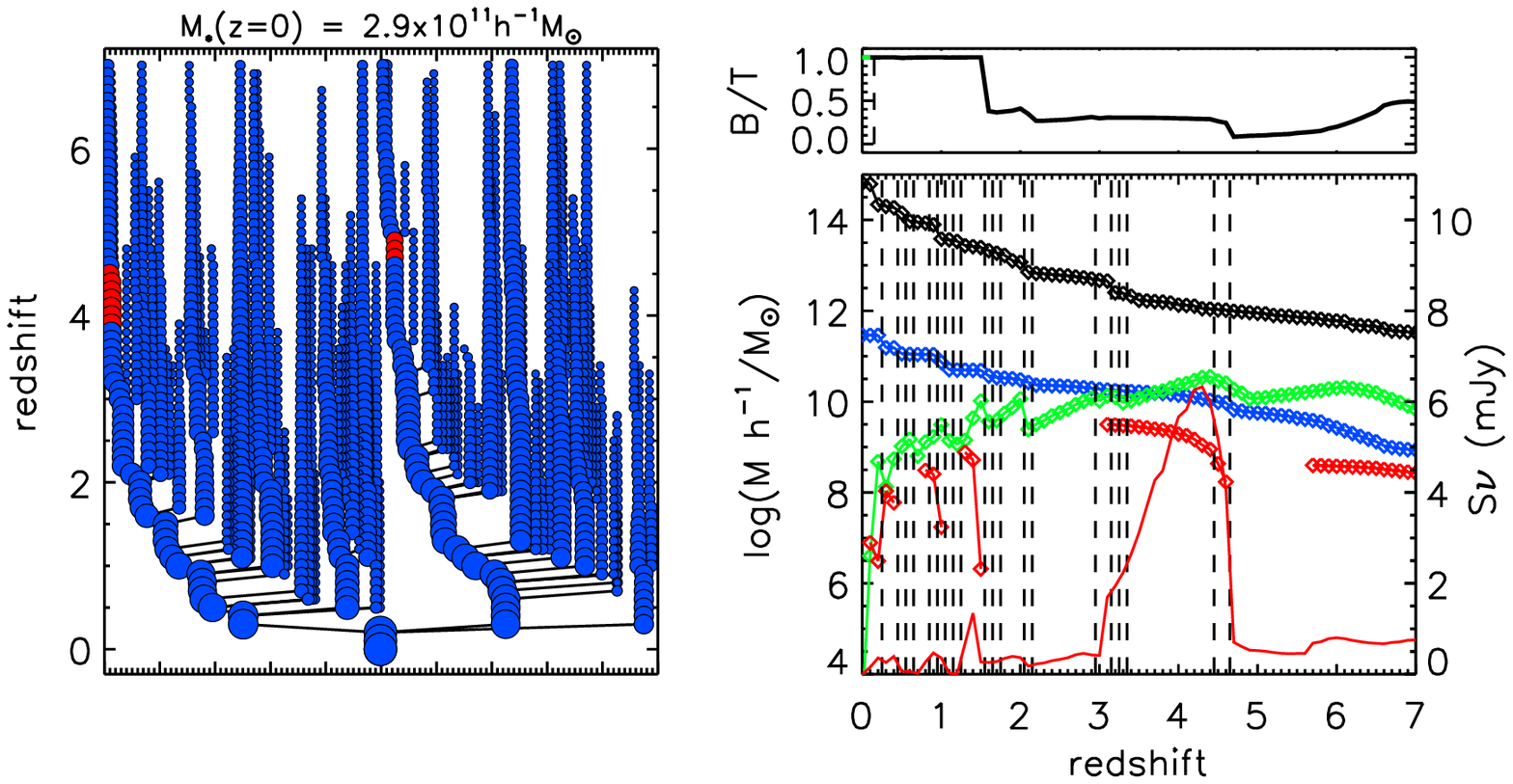}
\caption{ Left panels: galaxy merger trees for different $z=0$
galaxies. The size of the symbol is proportional to the stellar mass
as indicated by the key. The red circles indicate galaxies in a bright
SMG phase (when the $850\mu m$ flux exceeds 5 mJy). Right panels:
evolution with redshift of the galaxy on the main progenitor
branch. Top subpanels: evolution of the bulge-to-total stellar mass
ratio ($B/T$). The dashed vertical line indicates the redshift at
which the central galaxy (black line) becomes a satellite galaxy
(magenta line) after a halo merger. Lower subpanels: the symbols show
the evolution of the stellar mass (blue), cold gas mass (green), dark
matter host halo mass (black) and the mass formed in ongoing bursts
(red) in units shown on the left axis. The red line shows the
evolution of the $850\mu m$ flux, in units shown on the right axis,
and the dashed vertical black lines indicate the redshift at which the
galaxy undergoes a merger with another galaxy (see left panel).  From
top to bottom, the stellar masses of the galaxies at $z=0$ are
$M_{*}=2.7\times10^{10}, 1.0\times10^{11}, 2.9 \times10^{11}
\,h^{-1}\,M_{\odot}$ respectively. In the case of the upper two 
galaxy trees, we show all of the progenitors with stellar mass 
in excess of $10^{6} h^{-1}\,M_{\odot}$; in the lower tree, we plot 
progenitors with stellar mass greater than $10^{8} h^{-1}\, M_{\odot}$.
}
\label{GMT}
\end{figure*}

\subsection{Reprocessing of stellar radiation by dust}

The IR/sub-mm emission detected from SMGs is believed to originate
from the dust heated by radiation from young stars. Therefore a
treatment of absorption and reemission of radiation by dust is a key
element of the model. In \citet{Baugh05}, we calculated the
reprocessing of stellar radiation by dust using the {\tt GRASIL} code
\citep{Silva98,Granato00}, which calculates the distribution of dust
grain temperatures within each galaxy based on a radiative transfer
calculation and a detailed grain model. {The \grasil\ model has
been shown to be able to very successfully explain the SEDs of spiral,
irregular, elliptical and starburst galaxies, including ULIRGS, over a
huge range of wavelength, from the far-UV to the radio
\citep{Silva98,Vega08,Schurer09}. } However, a drawback of the {\tt
GRASIL} code is that it takes several minutes of CPU to run a single
galaxy, and so it is very time-consuming to run it on large samples of
galaxies. For the present paper, we need to follow the evolution
(including the sub-mm emission) of large samples of galaxies over many
timesteps, in order to be able select the small fraction of galaxies
undergoing an SMG phase at any cosmic time, and it is not
computationally feasible to do this using the {\tt GRASIL} code
directly.

We have therefore developed a simplified model of dust emission which
approximately reproduces the results from the {\tt GRASIL} code at
long wavelengths. {At the sub-mm wavelengths of interest here,
the SED of dust emission is insensitive to details of the dust model
such as the dust grain size distribution and the effects of small
grains and PAH molecules, allowing us to apply a simpler approach.} As
in {\tt GRASIL}, we assume that the dust is in two phases, molecular
clouds and a diffuse medium, with stars forming inside the molecular
clouds and then escaping into the diffuse medium on a timescale of a
few Myr. We use an approximate radiative transfer calculation
(different from the one in {\tt GRASIL} itself) to compute how much
stellar radiation is absorbed in each dust component. The most
important difference with {\tt GRASIL} is that in our simplified model
we assume the dust temperature within each of the dust phases is
uniform (so there are only two dust temperatures in a galaxy). In
contrast, in {\tt GRASIL} each size and composition of dust grain has
its own temperature, which depends on position within the galaxy
(which determines the intensity of the stellar radiation field which
heats it, via the radiative transfer calculation). Furthermore, {\tt
GRASIL} includes the effect of fluctuating grain temperatures for
small grains and PAH molecules. Thus in a full {\tt GRASIL}
calculation, there is a whole distribution of dust temperatures within
each galaxy. A final approximation we make in our simplified dust
model is to represent the dust emissivity by a power law in
wavelength. These differences between our two-temperature model and
{\tt GRASIL} are crucial when calculating the dust emission at mid-IR
wavelengths (which is dominated by emission from small grains and PAH
molecules, and by warm dust in star-forming clouds), but have much
less effect at the sub-mm wavelengths of interest in this paper. We
give more details about our simplified dust model in \citet{Lacey10b},
where we also show some comparisons with {\tt GRASIL}. In summary this
model works remarkably well at long wavelengths in the Rayleigh-Jeans
tail of the modified black body spectrum of the dust emission.

\section[]{Galaxy Merger Trees: Examples with SMG progenitors}
 
To follow galaxy evolution in {\tt GALFORM} we output the model
predictions at different redshifts. We label a galaxy as a bright SMG
when its $850\mu m$ flux exceeds 5~mJy. As examples of different
merger histories which can produce bright SMGs, we plot in the left
panels of Fig.~\ref{GMT}, galaxy merger trees for three different
present-day galaxies, covering a range in present stellar mass from
$\sim 3 \times 10^{10} \hMsol$ to $\sim 3 \times 10^{11} \hMsol$. In these
examples we require that a SMG phase appears in the main progenitor
branch (leftmost branch), which we define as follows: starting at 
the final time, we step back in time (up the tree), following the 
branch of the progenitor with larger stellar mass each time there 
is a merger. (Note that the main progenitor need not be the most
massive progenitor at all times.) We indicate with a red circle when a
galaxy at a given redshift is a bright SMG and with a blue circle
otherwise. In these selected examples, the more massive present-day
galaxies can show more than one bright SMG episode.

In the right panels of Fig.~\ref{GMT} we show the evolution of the
$850\mu m$ flux for the main progenitor galaxy (leftmost branch) in
the trees. Following each merger that triggers a burst, there is a
rapid increase in the $850\mu m$ flux, followed by a slow decline.
The time interval when the flux is above 5~mJy is different in the
different examples. In the same panels, we also plot the evolution of
the stellar, gas and dark matter halo masses for the main progenitor,
together with the mass formed in ongoing bursts. In the upper
sub-panels, we plot the evolution of the bulge-to-total stellar mass
ratio $B/T$ as an indication of the morphology of the galaxy, and we
also indicate when the central galaxy became a satellite galaxy.

In the first example shown in Fig.~\ref{GMT}, the central galaxy
becomes a satellite galaxy at $z=2.3$ (note the large increase in the
host halo mass due to the merger with a more massive halo). In the
second example, the galaxy becomes a satellite at $z=0.3$ and in the
third example, the galaxy becomes a satellite at $z=0.1$. As
mentioned previously, in the standard {\tt GALFORM} scheme, 
satellite galaxies cannot accrete any more
gas by cooling, so star formation can only continue until the
pre-existing cold gas reservoir is used up, which limits the increase
in stellar mass once a galaxy becomes a satellite (green line).

We also indicate in the right-hand panels when the galaxy undergoes a
merger with another galaxy. If there is a burst (see Section~2.3) the
gas from the merging pair is used up to produce a burst of star
formation (which raises the $850\mu m$ flux). Note that the galaxy
mergers at $z=2.7$ and $z=2.5$ in the first and second examples,
respectively, are minor mergers since the morphology of the resulting
galaxy is a mix of bulge+disk ($B/T \sim 0.5$ in both cases). The
galaxy mergers at $z=5.9$ and $z=2.1$ in the first and second
examples, respectively, are major mergers since the resulting galaxy
is a pure bulge galaxy with $B/T=1$. Since more hot gas can be
accreted by cooling onto central galaxies, more star formation takes
place, changing again the morphology of the galaxy. In the third 
example, the main progenitor becomes bulge dominated following 
a major merger at $z \sim 1.5$ and remains bulge dominated, even though 
it is still the central galaxy in its halo until $z=0.1$. In this 
example, there are a number of mergers with other galaxies which add 
mass to the spheroid (indicated by the vertical dashed lines in the main 
panel on the bottom right).

\section{Properties of SMGs at different redshifts}

Having identified all the SMGs in the model, we can study the
evolution of their properties.  In the model, the redshift
distribution of SMGs shows a peak at $z \sim 2$ (\citeauthor{Baugh05},
see also \citeauthor{Swinbank08}) similar to that observed by
\citet{Chapman03,Chapman05}.  For convenience in what follows, we define two
types of submillimetre galaxies according to a threshold in the
$850\mu m$ flux. We define a galaxy as a {\em bright} SMG when its
flux exceeds 5mJy and as a {\em faint} SMG when its flux exceeds 1mJy.
The faint sample therefore includes the bright sample, but these
bright galaxies only account for a small fraction of the galaxies in
the faint sample.  Most currently observed SMGs have fluxes around
5mJy or brighter (and so are bright SMGs by the definition
used here).  We present the faint SMGs as a comparison sample.

Our samples of SMGs are effectively volume limited. We consider dark
matter halo merger histories for a grid of halo masses, generating
many different examples for each mass on the grid.  SMGs are
identified according to their flux as outlined above, given the
redshift of a particular branch in the merger history.  The
distributions of properties from different mass trees are combined by
assigning a weight to each tree based on the abundance of haloes of
that mass and the number of example histories produced.

\subsection{Duration of the SMG phase }

\begin{figure}
\includegraphics[width=8.6cm, bb=108 379 541 705]{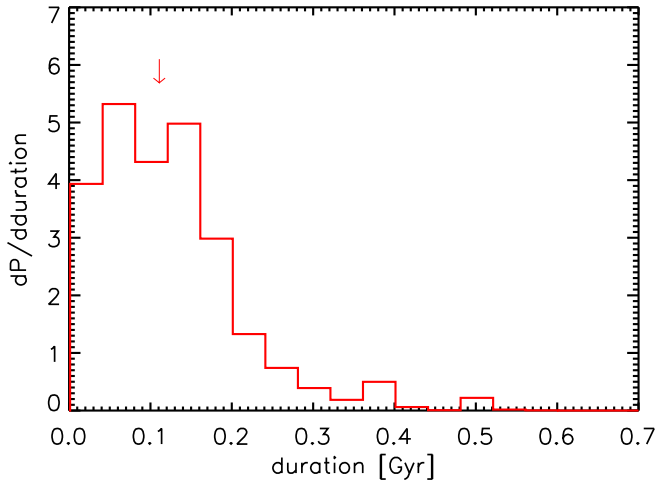}
\caption{ Duration of the bright SMG phase, i.e. the time for which
the flux is $S_{\nu}(850\mu m) > 5 {\rm mJy}$, for SMGs at
$2<z<3$. The distribution is that for a volume-limited sample. The
median of the distribution is 0.11Gyr (indicated by the arrow). The
distribution is normalized to unit area. }
\label{SMGtime}
\end{figure}

We define the duration of the bright SMG phase to be the time for
which the $850\mu m$ flux of a galaxy is above 5~mJy. This time is
typically smaller than the duration of the star formation burst which
triggered the flux increase (see the examples in Fig.~\ref{GMT}). In
Fig.~\ref{SMGtime} we plot the distribution of SMG phase durations all
of the bright SMG phases identified in the redshift range $2<z<3$. The
distribution is that for a volume-limited sample. The durations range
from $< 0.04$~Gyr to {roughly 0.5~Gyr}. The median of the
distribution is 0.11~Gyr. For comparison, the median duration of the
star formation bursts that triggered the bright SMG phase is
0.66~Gyr. {These predicted durations are similar to
observational estimates based on UV spectral fitting \citep{Smail03},
duty cycle arguments \citep{Chapman05} and gas depeletion timescales
\citep{Tacconi06}.}

\begin{figure*}
 \includegraphics[width=14cm]{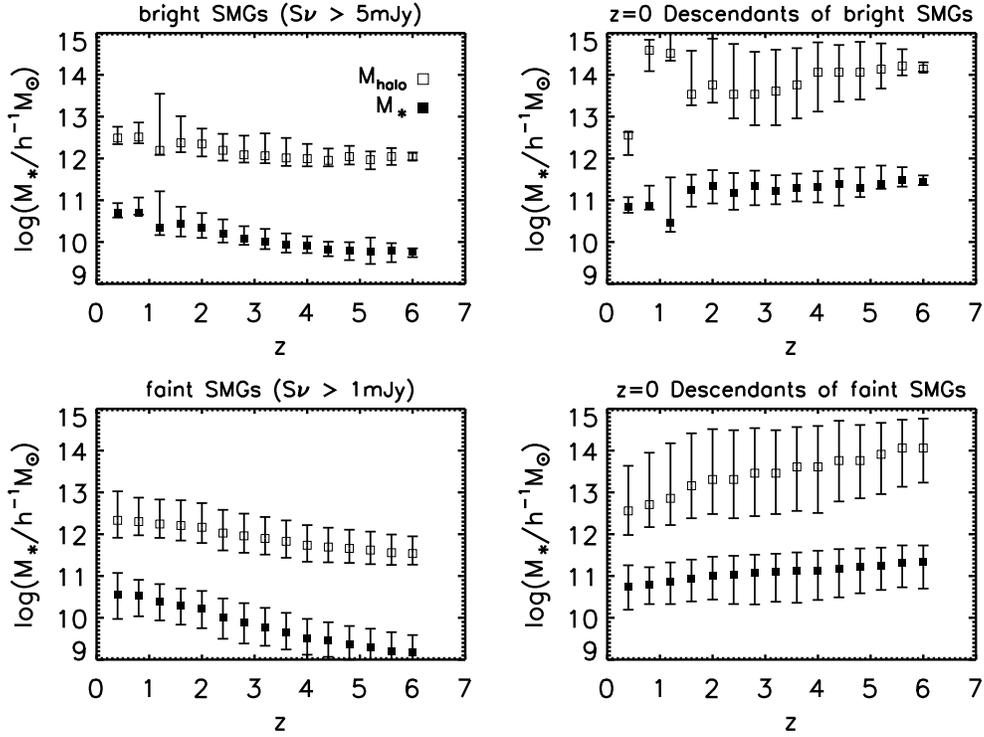}
\caption{Left panels:
The median stellar and host halo masses for SMGs selected at different
redshifts. Right panels: median stellar and host halo masses of the
z=0 descendants of SMGs identified at different redshifts. Top panels:
bright SMGs selected with $S_{\nu} > 5 {\rm mJy}$. Bottom panels:
faint SMGs selected with $S_{\nu} > 1 {\rm mJy}$. Filled squares show
stellar masses and open squares show halo masses. Errorbars indicate
the 10\% and 90\% percentiles. The medians are all calculated for
volume-limited samples of SMGs at each redshift.}
\label{Mstarevol}
\end{figure*}

\begin{figure*}
 \includegraphics[width=14cm]{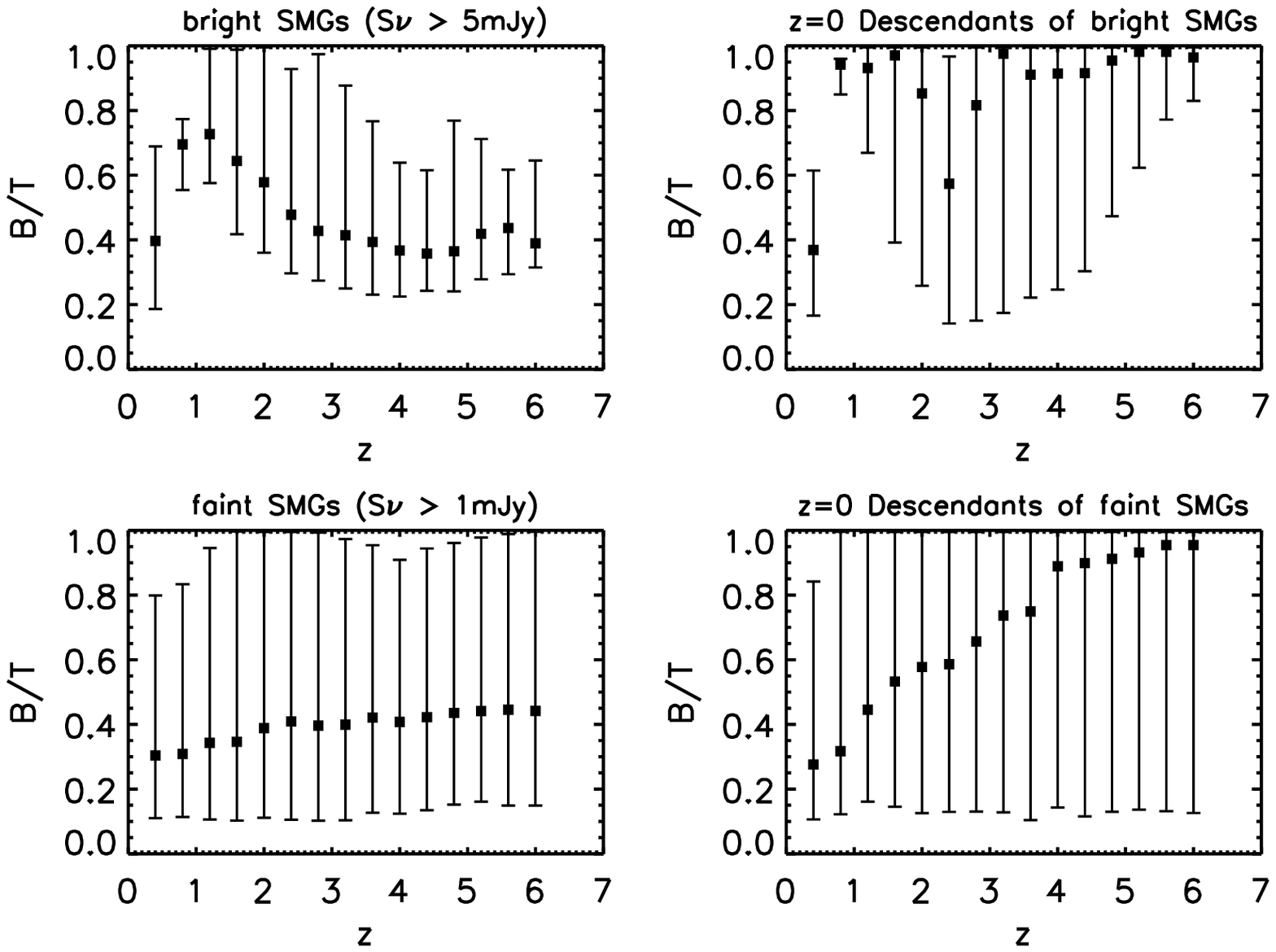}
\caption{Left panels: The predicted distribution of bulge-to-total
stellar mass ratio $B/T$ for SMGs selected at different
redshifts. Right panels: the $B/T$ distribution of the z=0 descendants
of SMGs identified at different redshifts. Top panels: bright SMGs
selected with $S_{\nu} > 5 {\rm mJy}$. Bottom panels: faint SMGs
selected with $S_{\nu} > 1 {\rm mJy}$. The filled-squares show the
median of the distribution at each redshift and the errorbars indicate
the 10\% and 90\% percentiles. The medians are all calculated for
volume-limited samples of SMGs at each redshift.}
\label{BTevol}
\end{figure*}

\subsection{Stellar and Halo Masses}

The left-hand panels of Fig.~\ref{Mstarevol} show the median stellar
and halo masses of SMGs identified at different redshifts (the
right-hand panels of this plot show the properties of the descendant
galaxies and are discussed in the next section).  The medians are
calculated for volume-limited samples of SMGs at each redshift. The
top left panel of Fig.~\ref{Mstarevol} shows the median stellar mass
and host halo mass for the bright SMGs as a function of redshift. For
the bright SMG sample at redshift z=1.5, the stellar and parent halo
mass are respectively {4 and 2} times more massive than the
bright SMGs identified at redshift z=6.  For comparison, we show also
the results for faint SMGs (i.e. those with $S_{\nu} > 1 {\rm mJy}$)
in the bottom left panel. Faint SMGs identified at redshift z=6 have
typically $1/4$ of the stellar mass of the bright SMGs identified at
the same redshift, and they live in haloes with $1/3$ the
mass. {By redshift z=1.5, this difference in host masses of
faint relative to bright SMGs has shrunk to around a factor $0.7$ for
both stellar and halo masses.}

{At redshift $z= 2$, close to the median redshift of bright
($S_{\nu}>5\mJy$) SMGs (both in the model and in the observational
data), the model predicts median stellar masses of $2.1\times 10^{10}$
and $1.7\times 10^{10}\hMsol$ for the bright and faint SMGs, and host
halo masses of $2.2\times 10^{12}$ and $1.5\times 10^{12}\hMsol$
respectively.}  The predicted stellar and halo masses of SMGs in this
model were compared with observational constraints in
\citet{Swinbank08}. The main observational constraint on halo masses
comes from measurements of galaxy clustering. \citeauthor{Swinbank08}
found good agreement between preliminary predictions of SMG clustering
from the model and observational data, but this issue will be examined
in more detail in \citet{Almeida10b}. \citeauthor{Swinbank08} also
found that the predicted stellar masses of SMGs were lower than
observational estimates based on fitting stellar population models to
broad-band fluxes.  However, this comparison is complicated by the
fact that the model includes a top-heavy IMF in starbursts, while the
observational estimates assume a solar neighbourhood IMF for all stars
(see also the discussion in \citet{Lacey10a}). {A more recent
observational analysis by \citet{Hainline10}, using updated stellar
population models and correcting for AGN contamination of the
rest-frame near-IR light, finds a median stellar mass of $\sim 5\times
10^{10}\hMsol$, for a sample of SMGs with $S_{\nu}\sim 5\mJy$ and
$z\sim 2$, assuming a \citet{Kroupa01} IMF. (Note that the typical
stellar masses found by \citeauthor{Hainline10} are significantly
lower than the values estimated by \citet{Michalowski10} for a similar
SMG sample, but not allowing for AGN contamination of the broad-band
SED.) The median stellar mass found by \citeauthor{Hainline10} is not
very different from the median value predicted by our model for bright
SMGs at $z=2$, even before allowing for the difference in IMFs and
star formation histories between the model and what is assumed in the
observational analyses.}  We will investigate this issue in more
detail in a future paper.

\subsection{Morphology}

In a similar way, we can study the morphologies of SMGs identified at
different redshifts. As mentioned above, a major merger produces a
pure bulge galaxy, whereas a minor merger can result in a galaxy with
a disk and a bulge (if the primary galaxy had a disk at the moment of
merging). {Note that in our model, the starburst responsible
for a galaxy appearing as an SMG starts {\em after} the two galaxies
have merged.}  The left-hand panels of Fig.~\ref{BTevol} show the
model predictions for the $B/T$ ratio of SMGs identified at different
redshifts, and the right-hand panels, which we discuss in the next
section, show $B/T$ for SMG descendants. As in Fig.~\ref{Mstarevol},
the medians are calculated for volume-limited samples of SMGs at each
redshift. In the top left panel of Fig.~\ref{BTevol}, we plot the
median of the bulge-to-total stellar mass ratio $B/T$ (where $B/T$=0
indicates a pure disk galaxy and $B/T$=1 indicates a pure bulge
galaxy) for the bright SMGs. At high redshift, {bright SMGs
have intermediate values of $B/T$ ($B/T \sim 0.4$), but the typical
$B/T$ increases to $\sim 0.7$ (indicating a bulge-dominated galaxy) as
the redshift decreases to $z\sim 1$, before dropping again to $B/T
\sim 0.4$ at lower redshifts. At $z=2$, bright SMGs are predicted to
have a median $B/T=0.6$, indicating a midly bulge-dominated galaxy,
but with quite a broad distribution, from 0.4 to 1 (10-90\% range).}

For comparison, we show also the results for faint SMGs selected with
$S_{\nu} > 1 {\rm mJy}$ in the bottom left panel of
Fig.~\ref{BTevol}. These faint SMGs typically have intermediate values
of $B/T$ at all redshifts, {with a very wide range (from 0.1 to
1),} and the samples at higher redshifts have slightly higher values
of $B/T$ compared to those at lower redshift (showing an opposite
trend as compared to bright SMGs).

{\citet{Swinbank10} have rececently investigated the
morphologies of bright SMGs at $z\sim 2$ using HST optical and near-IR
imaging. They find that the light profiles are fit by S\'{e}rsic indices
$n \sim 2-2.5$, compared to the values $n=1$ and $n=4$ expected for
exponential disks or $r^{1/4}$-law spheroids respectively. This result
indicates that these SMGs are mildy bulge-dominated, entirely
consistent with the $B/T$ values predicted by our model for SMGs at
similar fluxes and redshifts.}

\subsection{The SMG phase trigger: minor or major mergers?}

\begin{figure}
\includegraphics[width=8.6cm, bb=88 379 541 705]{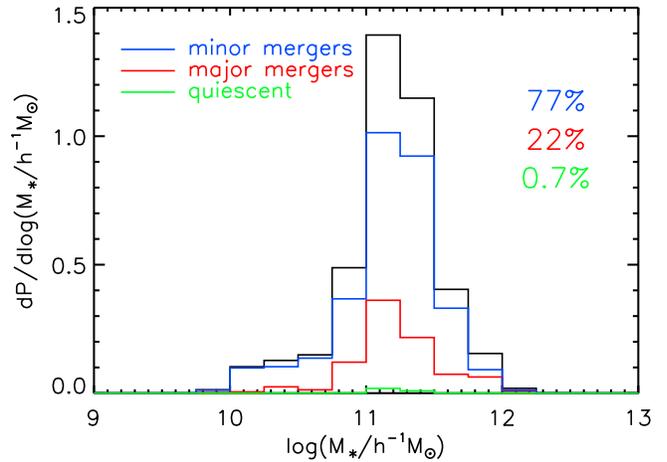}
\caption{ The distribution of stellar masses of the descendants of
  bright ($S_{\nu} > 5 {\rm mJy}$) SMGs at $z>1$, showing the separate
  contributions of SMGs produced by major mergers (red), minor mergers
  (blue) and quiescent galaxies (green). The contributions to this
  distribution from different redshifts have been weighted so as to
  correspond to what would be seen in a flux-limited survey over a
  fixed solid angle. The total distribution is normalized to have unit
  area.  }
\label{MinMajMergers}
\end{figure}

As we discussed before, most ($>99\%$) of bright SMGs in our model are
starbursts triggered by galaxy mergers. Bursts happen in all major
mergers and in some minor mergers (see Section~2.3). Which type of
merger dominates the triggering of SMGs? We answer this question in
Fig.~\ref{MinMajMergers}, where we plot the distribution of stellar
masses of $z=0$ descendants of bright SMGs at $z>1$ separated into
those triggered by minor mergers (blue) and produced by major mergers
(red). In this plot, we have weighted the contributions to the
distribution from the different redshift intervals by the comoving
volume per solid angle, so that the final distribution (which is
normalized to unit area under the histogram) corresponds to what would
be seen in a flux-limited sample of SMGs over a fixed solid angle.  We
find that $77\%$ are produced by minor mergers, $22\%$ are produced by
major mergers, and $0.7\%$ are quiescent galaxies (i.e. not ongoing
starbursts).  Minor and major mergers are responsible for similar
numbers of SMGs for the highest decendant masses, but for all other
descendant masses, minor mergers predominate. These proportions are
qualitatively consistent with the result shown in Fig.~\ref{BTevol}
that the bulge-to-total stellar mass ratios $B/T$ of SMGs are
typically intermediate between pure disk and pure bulge.

\begin{figure*}
 \includegraphics[width=14cm]{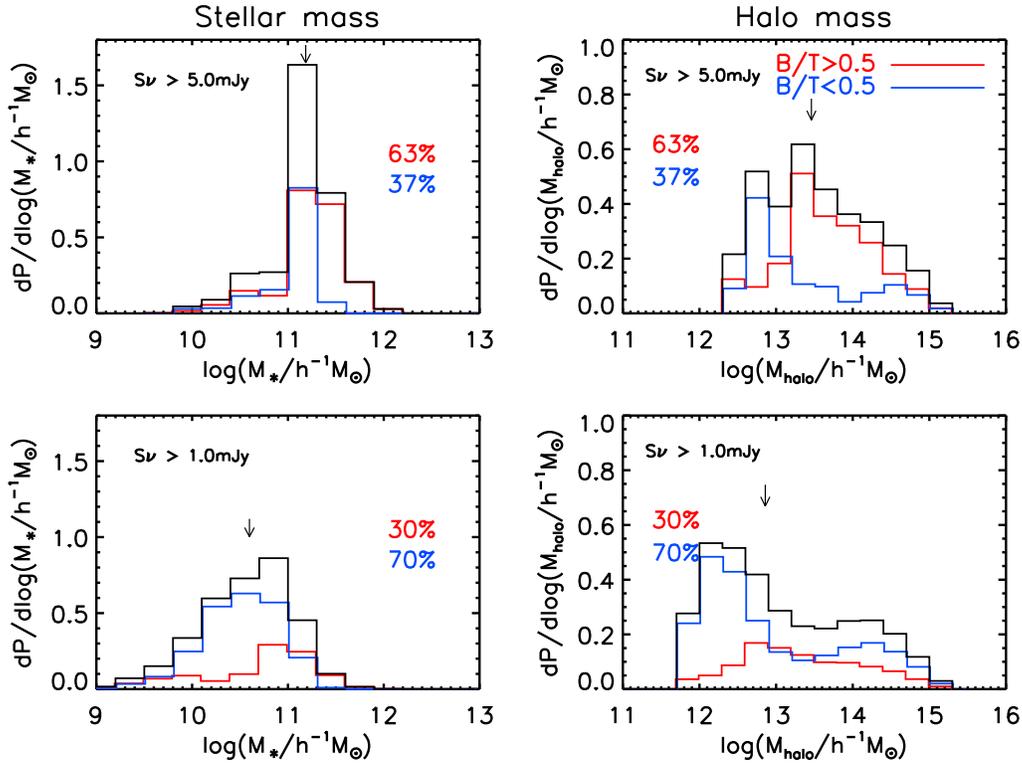}
\caption{Stellar mass distribution (left panels) and halo mass
distribution (right panels) for $z=0$ descendants of bright $z>1$ SMGs
($S_{\nu} >5 {\rm mJy}$, top panels) and for descendants of faint
$z>1$ SMGs ($S_{\nu} >1 {\rm mJy}$, bottom panels). The distributions
are for volume-limited samples of descendants, and weight each
descendant equally whether it had one or more SMG progenitors. The red
distribution is for bulge-dominated ($B/T>0.5$) descendant galaxies
and the blue for disk-dominated ($B/T<0.5$) descendant galaxies. The
black lines represent the sum of the red and blue distributions. The
mass distributions are normalized to unit area for the full
sample. The vertical black arrows indicate the median of the total
distribution in each panel.  The percentages in each panel refer to
the percentage of bulge-dominated (in red) and disk-dominated (in
blue) galaxies.  }
\label{mstdescendantsBT}
\end{figure*}

\begin{figure*}
 \includegraphics[width=14cm]{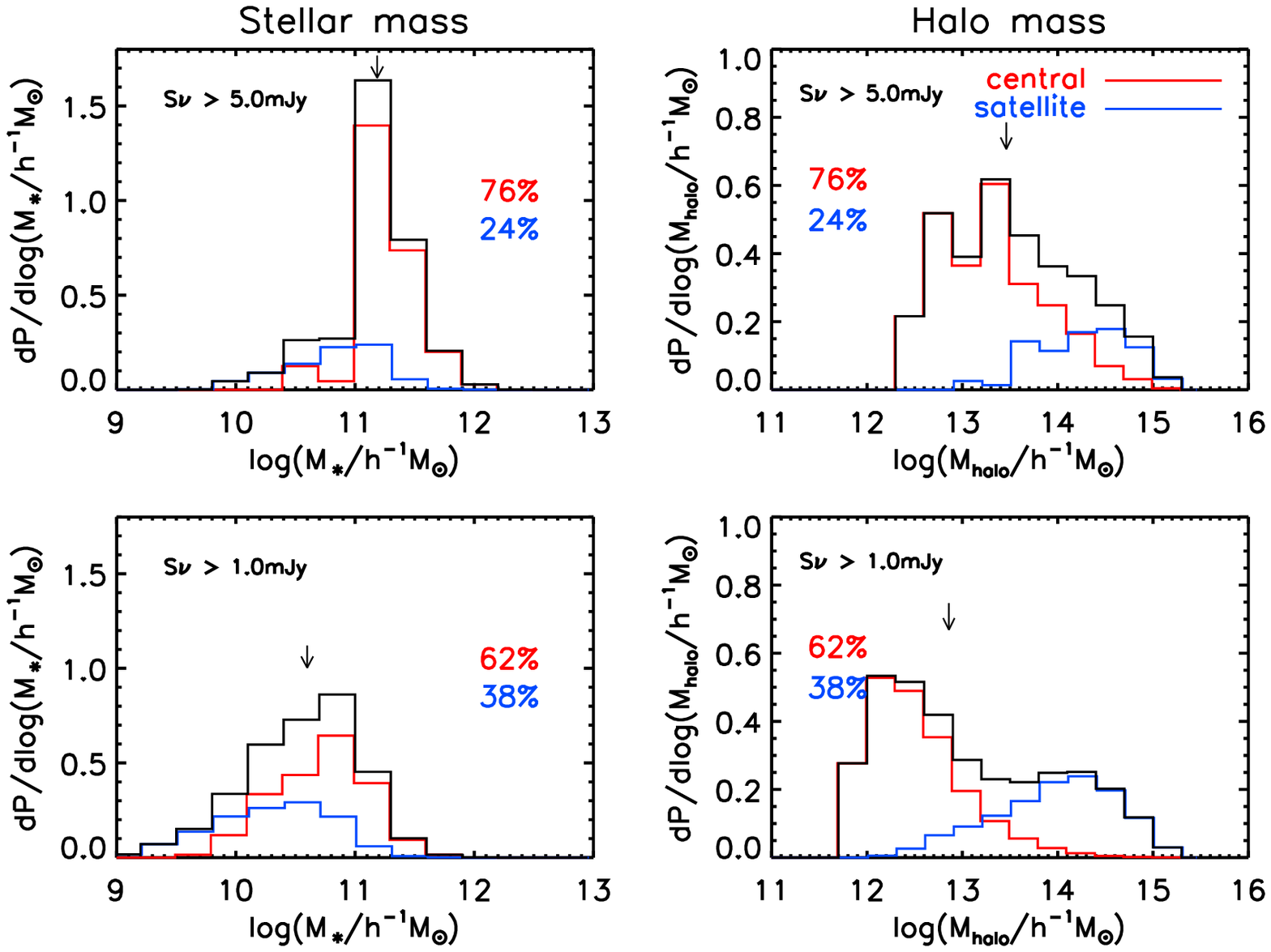}
\caption{Similar to Fig.\ref{mstdescendantsBT} but separating the
  descendant galaxies into central (red) and satellite (blue)
  galaxies.  In this case, the percentages in each panel refer to the
  percentage of central (upper value) and satellite (lower value)
  galaxies.  }
\label{mstdescendantscentral}
\end{figure*}

\section{Properties of SMG descendants }

Most of the SMGs detected in flux-limited surveys lie at redshifts
z=1-4.  SMGs at redshift $z<1$ account for only a small fraction of
the population in a flux limited sample due to the relatively 
small volume at low redshift. Model SMGs at low redshifts $z<1$ are 
mostly objects with modest star formation rates which have bright sub-mm fluxes 
simply due to their proximity to us compared with the bulk of the SMG population, 
rather than because an intense episode of star formation has taken place. 
In this section, we will focus on the descendants of the {\em high-z} SMGs, 
which we define to be those with $z>1$ on the basis that these have the largest 
overlap with observational samples, and we will investigate the distribution 
in stellar and parent halo mass, morphology and central/satellite galaxy 
classification.

\subsection{Stellar and host halo masses of SMG descendants}
\label{ssec:mass_desc}

What are the masses of the present-day descendants of SMGs seen at
different redshifts?  In the right panels of Fig. 3, we plot the
median stellar and host halo masses of the z=0 descendants of the SMGs
identified at different redshifts. In these panels, the medians are
calculated for a volume-limited sample of SMG progenitors, so that
descendants are weighted according to the number of such
progenitors. Note that it is possible for a present-day galaxy to have
multiple SMG progenitors in different branches of its merger tree (see
Fig.~\ref{GMT}).  We find that bright SMGs at higher redshift
typically evolve into somewhat more massive galaxies hosted in more
massive haloes compared to bright SMGs found at lower redshifts. The
model predicts that galaxies seen as bright SMGs at the median of the
observed redshift distribution ($z \sim 2$) increase their stellar
mass by one order of magnitude by redshift z=0 ( from $\sim 2 \times
10^{10}h^{-1}M_{\odot}$ to $\sim 2 \times 10^{11}h^{-1}M_{\odot}$).

\begin{figure}
 \includegraphics[width=8.6cm, bb=100 100 550 705]{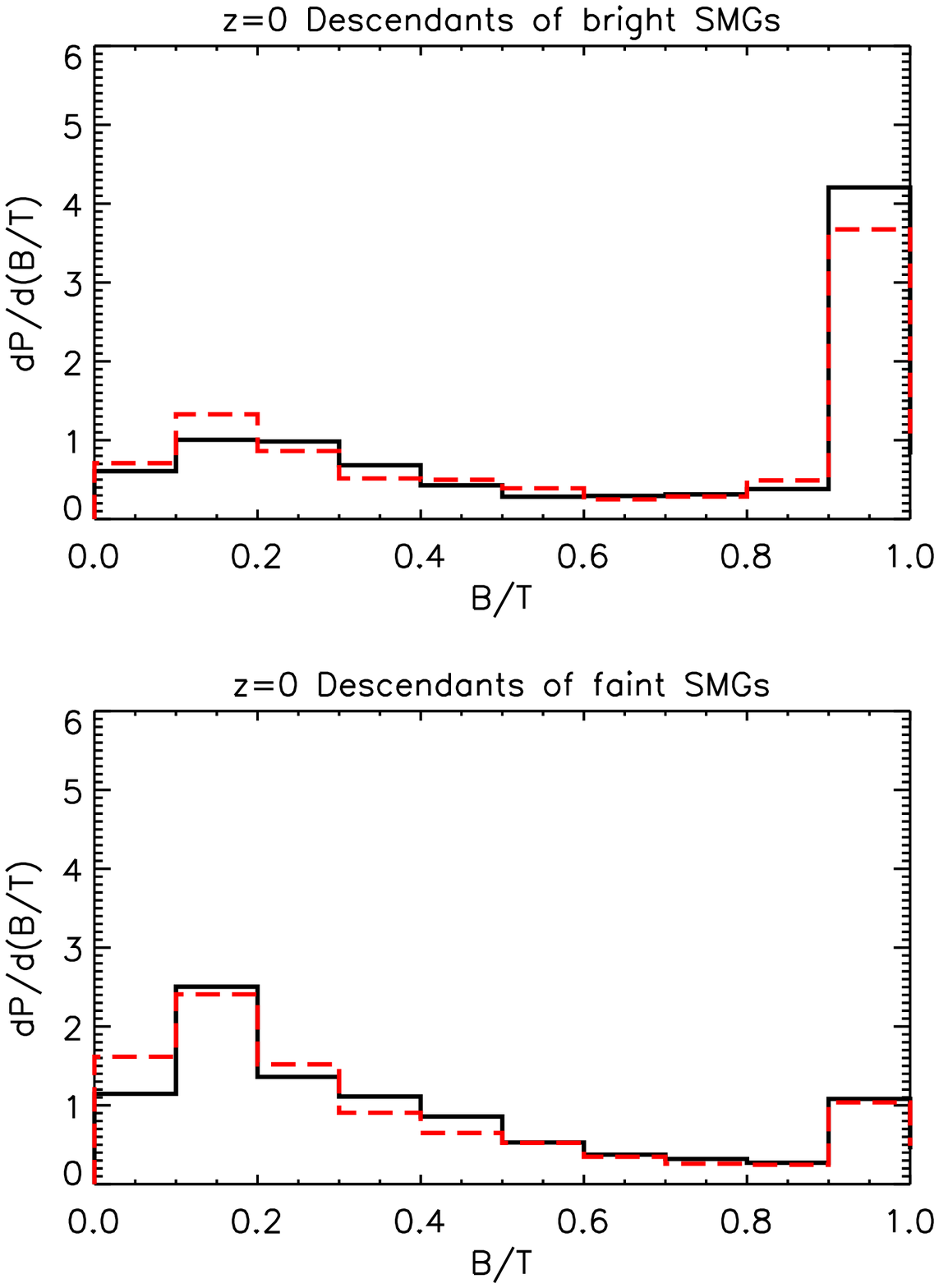}
\caption{Probability distribution of the bulge-to-total stellar mass
ratio ($B/T$) for $z=0$ descendants of bright $z>1$ SMGs (top panel,
black) and for descendants of faint $z>1$ SMGs (bottom panel,
black). The distributions are for volume-limited samples of
descendants, and give each descendant the same weight whether it had
one or more SMG progenitors. The dashed red lines show the $B/T$
distributions of the comparison samples formed by galaxies at $z=0$
with a similar stellar mass distribution to the bright and faint SMG
descendants. All distributions are normalized to unit area under the
histogram.}
\label{BTdescendants}
\end{figure}

In Fig.~\ref{mstdescendantsBT} we show the distributions of stellar
and halo mass for the z=0 descendants of galaxies which had SMG
progenitors at $z>1$. In this figure, and also in
Figs.~\ref{mstdescendantscentral} and \ref{BTdescendants}, the
distributions are calculated for volume-limited samples of SMG
descendants, with each descendant given the same weight whether it had
one or more SMG progenitors.  The top left panel of
Fig.~\ref{mstdescendantsBT} shows the distribution of stellar masses
of the descendants of all bright SMGs with $z>1$.  We separate the
galaxies into bulge-dominated ($B/T >0.5$) and disk-dominated ($B/T <
0.5$) samples. The number of bulge-dominated descendants is nearly
double the number of disk-dominated descendants. Also, the
bulge-dominated descendant galaxies have a higher median mass. For all
of the descendants, the median stellar mass is $1.5 \times
10^{11}h^{-1}M_{\odot}$.  The distribution of host dark matter halo masses of
the z=0 descendants, plotted in the top right panel of
Fig.~\ref{mstdescendantsBT}, shows a higher median mass for
bulge-dominated than for disk-dominated descendants. The median dark
matter halo mass for all $z=0$ descendants of bright $z>1$ SMGs is
$2.9 \times 10^{13}h^{-1}M_{\odot}$.
For comparison, we also show the results for the faint SMGs selected
with $S_{\nu} > 1 {\rm mJy}$. The median stellar and dark matter halo
masses of the descendants are about 4 times smaller than for the bright
SMGs, and the distributions are wider. In this case, the proportions of
disk-dominated and bulge-dominated descendants are reversed compared
to the bright SMGs, with more than twice as many disk-dominated as
bulge-dominated descendants.

In Fig.~\ref{mstdescendantscentral}, we repeat the same analysis but
separate the descendants into central and satellite galaxies in the
halo where they live at the present-day. We can see that nearly 80\%
of the $z=0$ descendants of bright SMGs are central galaxies, while
this fraction is reduced to 60\% for the faint SMGs. We note that
satellite galaxies that are descendants of bright SMGs are found only
in the more massive haloes today.

{We note that \citet{Hainline10} arrived at a similar estimate
  to our own for the masses of the present-day descendants of bright
  SMGs, but starting from observations of SMGs at high redshift, and
  making different assumptions from ours about their subsequent
  evolution down to $z=0$.}

\subsection{Morphology of descendants of SMGs compared to other similar mass
  galaxies}

The top right panel of Fig.~\ref{BTevol} shows us that the model
predicts that bright SMGs identified at different redshifts typically
evolve into highly bulge-dominated galaxies at redshift z=0.
According to the bottom right panel of Fig.~\ref{BTevol}, the $z=0$
descendants of faint SMGs at low and intermediate redshifts have
intermediate values of $B/T$. Only the higher redshift samples evolve
predominantly into highly bulge-dominated systems at the present-day.

Fig.~\ref{mstdescendantsBT} shows that the majority of the descendants
of bright $z>1$ SMGs are bulge-dominated systems, while for faint SMGs
most descendants are disk-dominated. In Fig.~\ref{BTdescendants} we
show (in black) the probability distribution of $B/T$ for the $z=0$
descendant galaxies of $z>1$ SMGs (with each descendant given equal
weight whether it had one or more SMG progenitors). The top panel
shows the descendants of bright SMGs, of which $63\%$ of the
descendants are bulge-dominated ($B/T > 0.5$) and $37\%$ are
disk-dominated ($B/T < 0.5$). Among these, $42\%$ of the descendants
are pure bulge systems ($B/T > 0.9$) and only $6\%$ are pure disk
systems ($B/T < 0.1$). The bottom panel of Fig.~\ref{BTdescendants}
shows the probability distribution of $B/T$ for the descendants of
faint SMGs. In this case, only $30\%$ of the descendants are
bulge-dominated systems, with 11\% being pure bulge systems, and 12\%
being pure disk systems.

However, are these results a special characteristic of SMG
descendants, or are these morphologies simply typical of $z=0$
galaxies with similar stellar mass? To answer this, we select
comparison samples (for bright and faint SMGs) composed of galaxies at
$z=0$ with similar stellar mass distributions to the bright and faint
SMG descendants respectively. In Fig.~\ref{BTdescendants}, we plot the
probability distribution of $B/T$ for the SMG descendants in black,
and for the comparison samples in red. The descendant and comparison
samples for bright and faint SMGs show very similar
distributions. Despite the fact that most of the descendants of bright
SMGs are bulge-dominated galaxies, their morphologies are not
significantly different from other $z=0$ galaxies with similar stellar
masses.

\begin{figure}
\begin{center}
 \includegraphics[width=7cm, bb=55 35 565 765, clip=true]{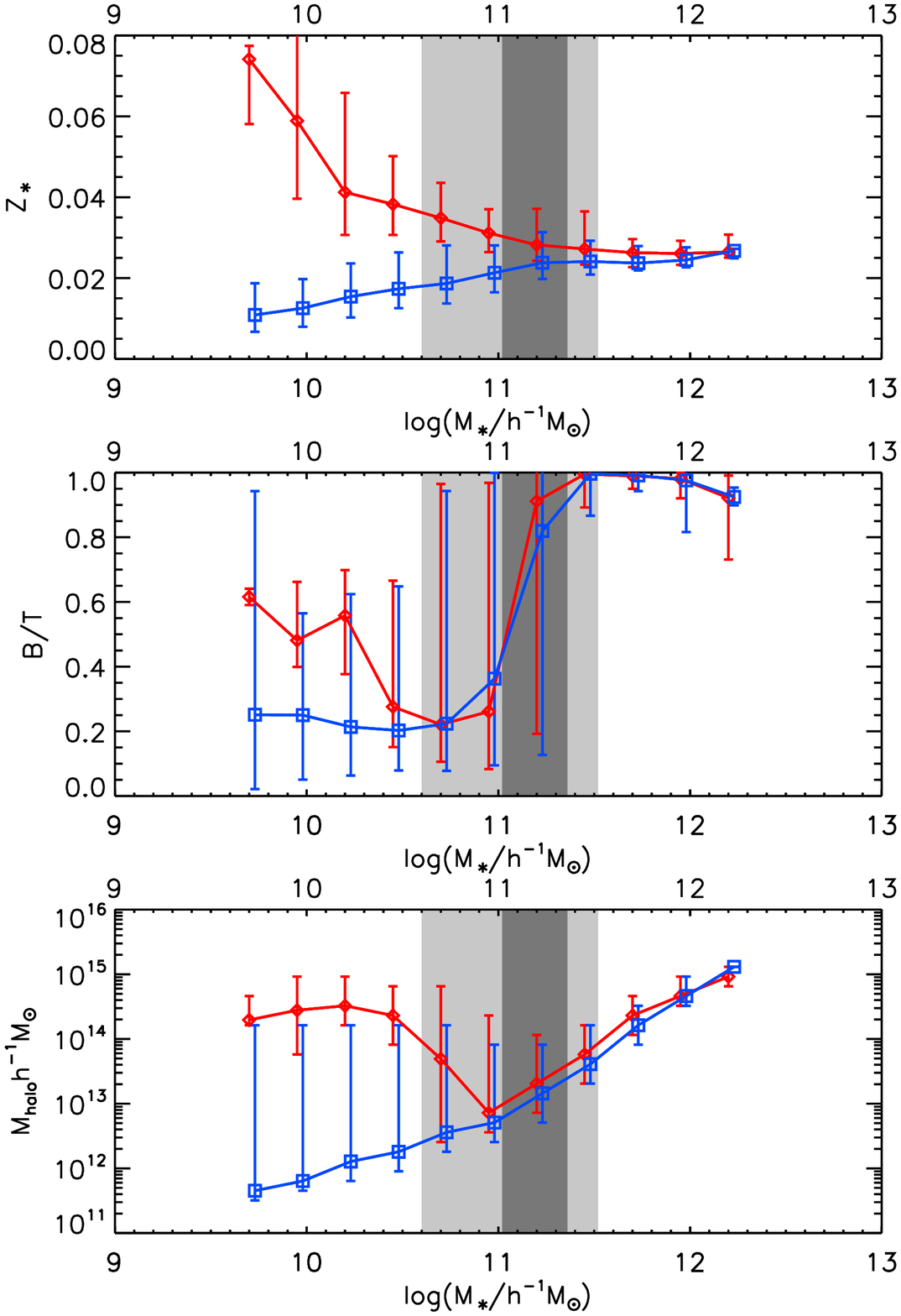}
\caption{ {The three panels show the median stellar
metallicity, bulge-to-total stellar mass ratio, and halo mass vs
stellar mass for present-day descendants (red line) and
non-descendants (blue line) of bright SMGs. The medians are calculated
for volume-limited samples of present-day galaxies. The error bars
show the 10-90\% range in bins of stellar mass. The dark grey shaded
region shows the 25-75\% range for the stellar masses of descendants
of bright SMGs, while the light grey shading shows the 10-90\% range.}
}
\label{Zdescendants}
\end{center}
\end{figure}

\subsection{Metallicity of descendants of SMGs compared to other
  galaxies}

{In the top panel of Fig.~\ref{Zdescendants}, we compare the
median stellar metallicity (weighted by stellar mass) as a function of
stellar mass for galaxies which are descendants of bright SMGs (red
curves) to that of other galaxies of the same stellar mass which are
not descendants of bright SMGs (blue curves). The medians are
calculated for volume-limited samples of present-day galaxies. The
dark and light grey shaded regions respectively show the 25-75\% and
10-90\% ranges for the stellar masses of descendants of bright
SMGs. We see that high stellar mass descendants of bright SMGs have
very similar metallicities to other present-day galaxies of the same
mass, while low stellar mass descendants typically have larger
metallicities than high-mass descendants or other galaxies of similar
masses. For descendants of the median stellar mass, the median
metallicity is only 20\% higher than for non-descendants of the same
mass, but for the 10\% least massive descendants, the difference in
median metallicity is a factor 2 or more. It thus appears that for
descendants below the median stellar mass, the starburst which powered
the SMG was also responsible for chemically enriching the descendant
galaxy above the typical level for galaxies of that present-day mass.}

{The lower two panels of Fig.~\ref{Zdescendants} make similar
comparisons of the bulge-to-total stellar mass ratio, $B/T$, and the
halo mass for descendants and non-descendants of bright SMGs. We see
that the most massive 90\% of descendants have very similar median
$B/T$ to non-descendants of the same mass, while the least massive
10\% have significantly larger $B/T$ than the non-descendants (up to a
factor 2-3). The halo masses for the most massive 75\% of descendants
are very similar to non-descendants of the same mass, but for the
least massive descendants, the median halo masses are much larger (up
to a factor $\sim 10^3$). }

\section{SMG progenitors of present-day galaxies}

\begin{figure*}
\includegraphics[width=14cm]{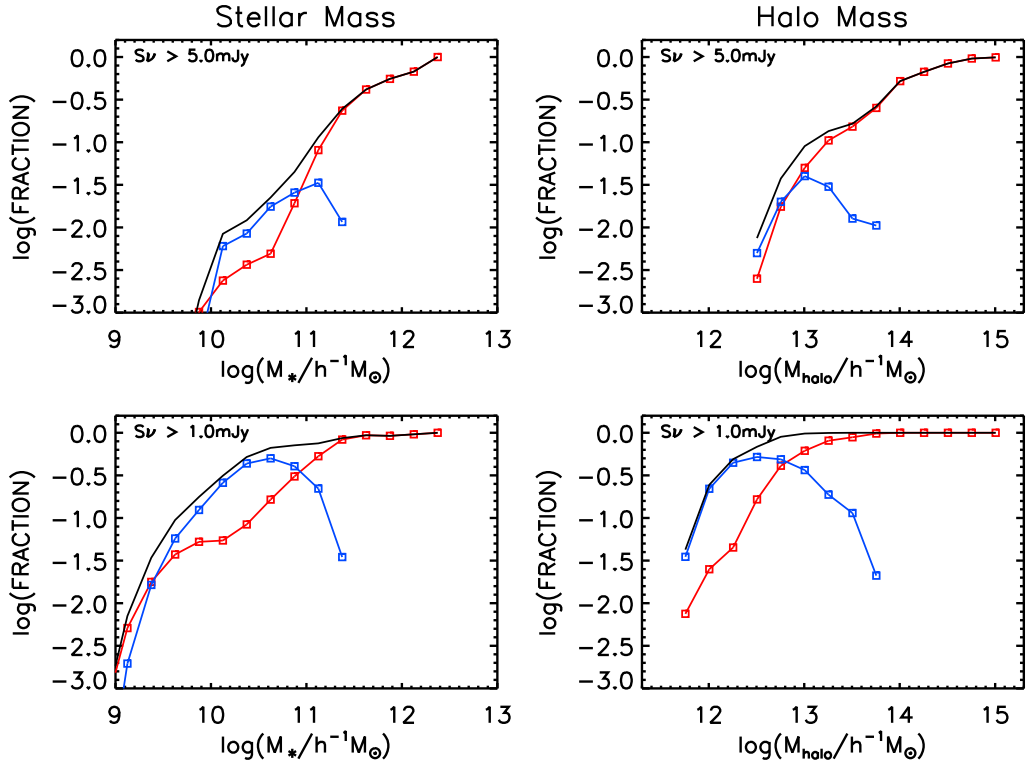}
\caption{Left panels:
fraction of present-day galaxies by stellar mass that are predicted to
be descendants of bright SMGs ($S_{\nu} >5 {\rm mJy}$, top panel) and
descendants of faint SMGs ($S_{\nu} >1 {\rm mJy}$, bottom panel).
Right panels: fraction of present-day halos by mass that are predicted
to have halo progenitors that hosted at least one bright SMG (top
panel) or at least one faint SMG (bottom panel). These are computed
for $z>1$ SMG progenitors. Left panels: bulge-dominated ($B/T>0.5$)
descendants in red and disk-dominated ($B/T<0.5$) descendants in
blue. Right panels: bulge-dominated ($B/T>0.5$) halo central galaxy in
red and disk-dominated ($B/T>0.5$) halo central galaxy in blue. The
black line represents all of the SMG descendants (the sum of the blue
and red lines).}
\label{mstarhaloz0}
\end{figure*}

In the previous section we studied the properties of the descendants
of SMGs. We identified SMGs in the model at a particular redshift and
we found their z=0 descendants. In this section we want to study the
connection of SMGs to present day galaxies from the opposite
perspective: given a galaxy today with a particular mass, what is the
probability that this galaxy had SMG progenitors? And for dark matter
haloes at present-day, what is the probability that a progenitor halo
hosted at least one SMG?  Fig.~\ref{mstarhaloz0} answers these
questions, for the $z=0$ descendants of both bright and faint SMGs at
$z>1$ (top and bottom panels respectively). The left panels show this
probability as a function of stellar mass, and the right panels as a
function of halo mass. We see that the probability that a galaxy or
halo had one or more SMG progenitors at $z>1$ brighter than a given
850$\mum$ flux increases with increasing mass. From the top left
panel, we see that present-day galaxies have a probability of 10\% to
be descendants of bright SMGs ($S_{\nu} > 5\mJy$) at a stellar mass of
$10^{11}h^{-1}M_{\odot}$, and 50\% at a mass of $6 \times
10^{11}h^{-1}M_{\odot}$. For comparison, we see from the lower-left
panel that 50\% of galaxies with present day stellar masses of $2
\times 10^{10}h^{-1}M_{\odot}$ are descendants of faint SMGs ($S_{\nu}
> 1\mJy$).  We find in both cases (bright and faint SMGs) that at
intermediate masses ($\sim 10^{10}-10^{11}\hMsol$) there are more
disk-dominated descendants of SMGs than bulge-dominated
descendants. For haloes, we see from the right panels that more than
$50\%$ of the haloes with masses larger than $10^{14}h^{-1}M_{\odot}$
are predicted to have bright SMG progenitors, and $50\%$ of the haloes
with masses larger than $2 \times 10^{12}h^{-1}M_{\odot}$ are
predicted to have faint SMG progenitors.

\section{Contribution of SMGs to the present-day stellar mass}

We saw in the previous section that most massive galaxies today are
predicted to be descendants of SMGs. We found also in Section 4.1 that
the median of the SMG phase duration is $0.11 {\rm Gyr}$. Since in the
literature a high star formation rate (SFR) is traditionally inferred
for the bright SMGs {(based on the assumption of a normal
  IMF)}, we want to know what contribution to the present-day stellar
mass is actually produced in the SMG phase.

\begin{figure}
 \includegraphics[width=8.6cm, bb=100 200 540 590]{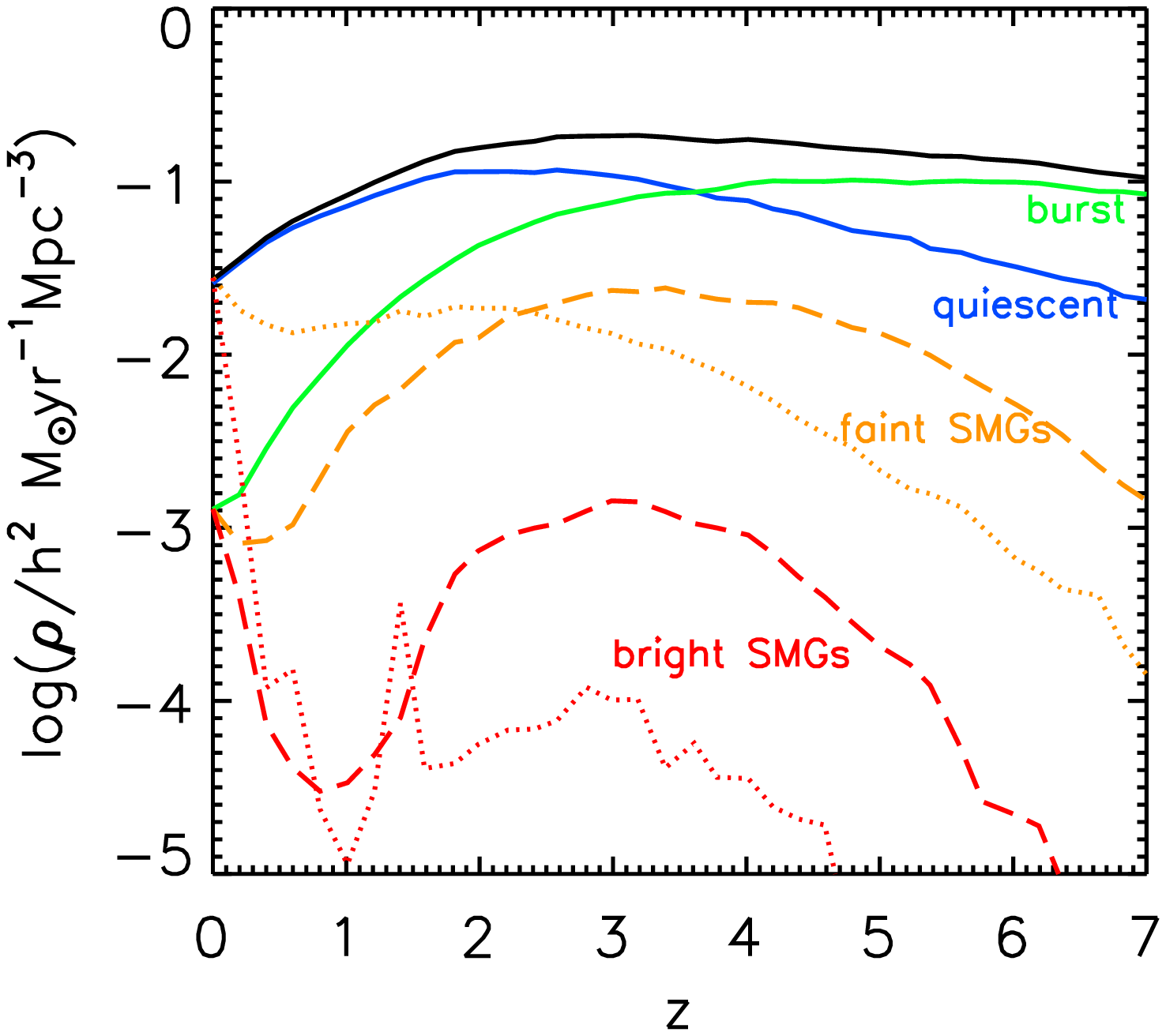}
\caption{Evolution of the cosmic star formation rate per unit comoving
volume. The black line shows the total total star formation rate,
while the green and blue lines show the separate contributions from
bursts and quiescent star formation respectively. The red and orange
lines respectively show the star formation rate in bright and faint
SMGs, for which the separate contributions from bursts and quiescent
star formation are shown by dashed and dotted lines, respectively.}
\label{sfrvsz}
\end{figure}

In Fig.~\ref{sfrvsz} we plot the predicted evolution of the cosmic SFR
density. We show both the total, as well as the contributions from
bursts and quiescent disks, and the contributions from galaxies in a
bright ($S_{\nu}>5\mJy$) or faint ($S_{\nu}>1\mJy$) SMG phase. We see
that the contribution to star formation from quiescent disks dominates
at redshift $z \lsim 3.5$, but bursts dominate at higher redshifts, as
found earlier by \citet{Baugh05}. At redshifts $z=2-4$, around $1\%$
of the cosmic star formation is produced by bursts in bright SMGs, and
around $10\%$ by bursts in faint SMGs.  The quiescent SFR density is
typically one order of magnitude smaller than the burst SFR density
for bright SMGs at high redshift. For faint SMGs, the burst SFR
density dominates at redshift $z\gsim 2$, but quiescent star formation
becomes more important at lower redshift. At low enough redshift
($z\ll 1$), star formation at any level in a galaxy is sufficient to
produce a sub-mm flux above the bright or faint thresholds, which is
why the quiescent and burst contributions from bright and faint SMGs
converge on the total quiescent and burst contributions at $z=0$.

\begin{figure}
\includegraphics[width=8.4cm, bb=80 100 540 700]{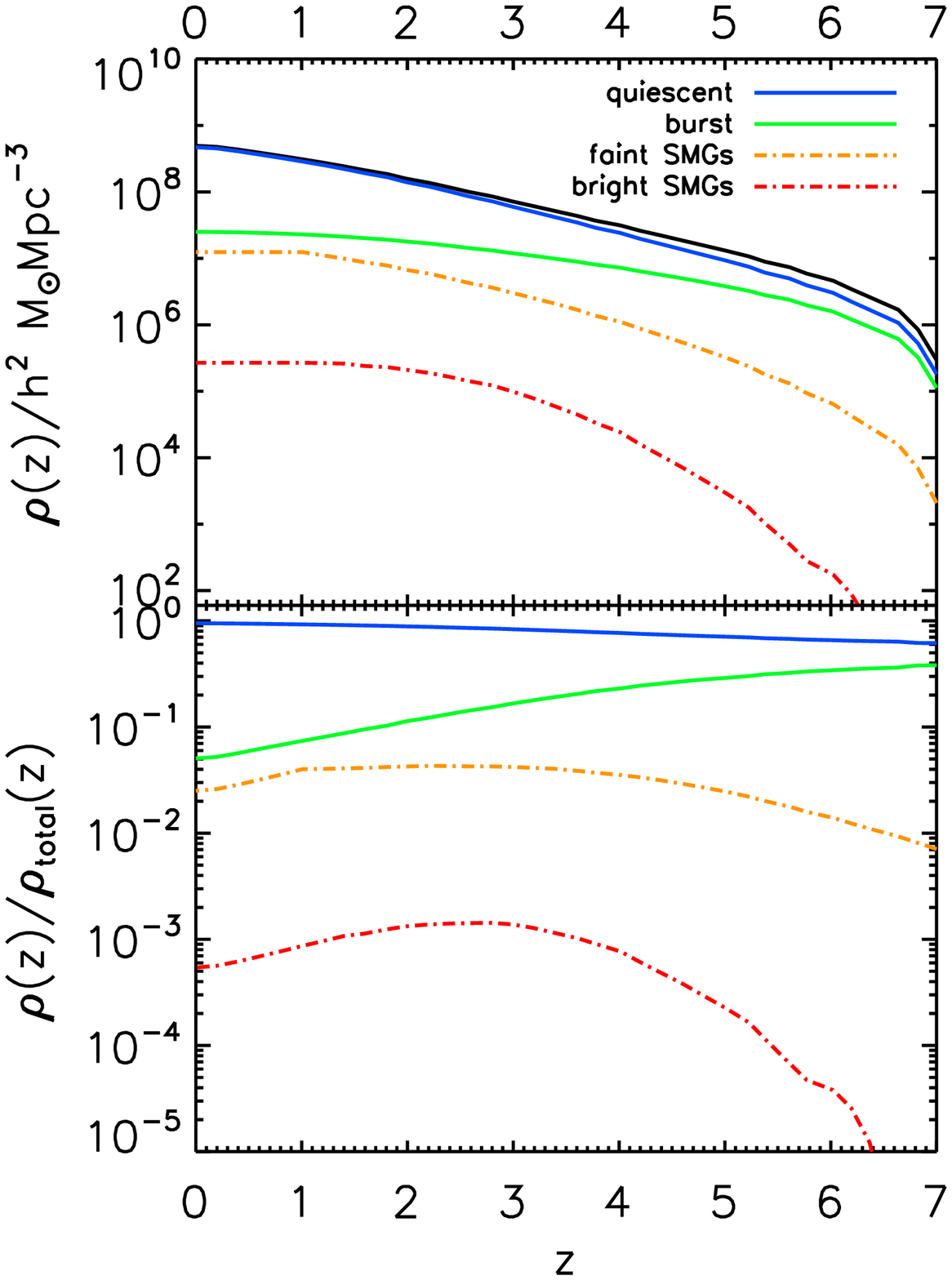}
\caption{Top panel: comoving stellar mass density as a function of
redshift. The black line shows the total, while the blue and green
lines show the contributions to this from quiescent and burst star
formation respectively, and the red and orange lines show the
contributions from star formation in bright or faint SMG phases. For
the SMGs, we only show the contributions from SMG phases at
$z>1$. Bottom panel: similar to the top, except that the different
contributions to the stellar mass density are plotted as fractions of
the total stellar mass density at that redshift.}
\label{cumsfr}
\end{figure}

In the top panel of Fig.~\ref{cumsfr} we plot the comoving stellar
mass density as a function of redshift, showing both the total, and
the contributions to this from star formation in the quiescent and
burst modes, and the contributions from star formation during bright
and faint SMG phases at $z>1$. Note that the stellar mass densities we
plot are the values after allowing for recycling of gas to the ISM
from dying stars, using the recycled fractions for the two IMFs given
in Sec.~\ref{ssec:IMF} (recall that the recycled fraction for the top
heavy IMF is close to unity). The lower panel shows the different
contributions to the stellar mass density plotted as fractions of the
total stellar mass density at that redshift. We see from this that
overall, bursts of star formation contribute a total of $5\%$ to the
present-day stellar mass density. Star formation (burst+quiescent) in
the faint SMG phase at $z>1$ contributes 2\% to the present-day
stellar mass density, while the bright SMG phase at $z>1$ contributes
only 0.06\%. {The contribution of bursts and SMGs to metal
production are however much larger, due to both the larger yield and
larger recycled fraction for the burst IMF compared to the quiescent
IMF. The model predicts that all bursts over the history of the
universe produced 70\% of the metals existing today, while bright SMGs
at $z>1$ produced 0.8\% of the metals existing today.}

\begin{figure}
 \includegraphics[width=8.6cm, bb=80 380 540 700]{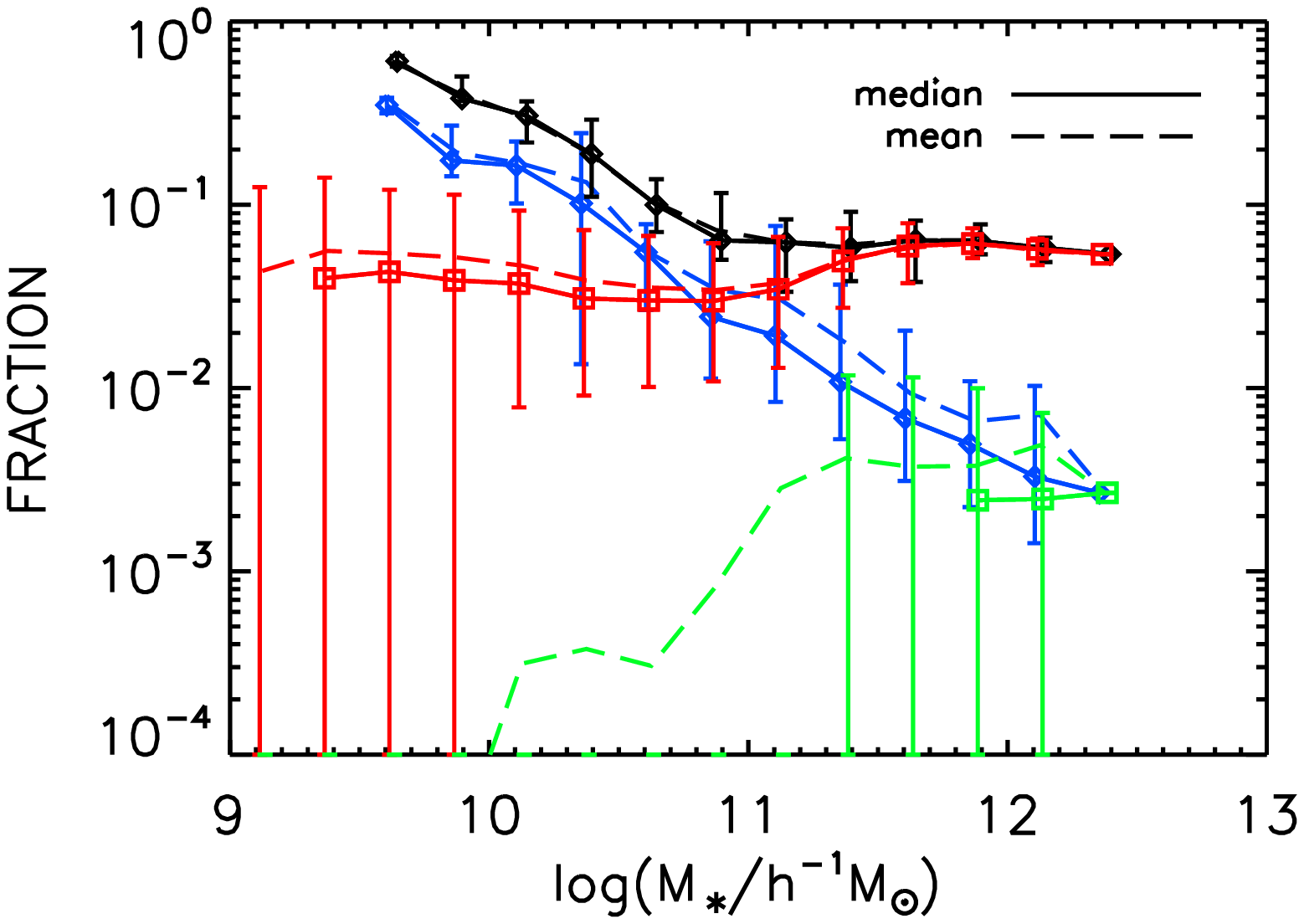}
\caption{The fraction of the stellar mass produced by different mechanisms as a
function of present-day stellar mass. The red lines show the average fraction
produced by all bursts in all galaxies, while the black lines show the
fraction produced by all bursts in descendants of bright SMGs
only. The green lines show the fraction produced by bursts and quiescent star
formation in the bright SMG phase for all galaxies, and the blue lines show the
fraction produced by bursts and quiescent star
formation during an SMG phase in descendants of
bright SMGs only. Note that in this plot we only include the
contribution of bright SMGs at $z>1$. The dashed lines show the mean, and
the solid lines the median, with the errorbars showing the 10\% and
90\% percentiles.}
\label{FracSMGmst}
\end{figure}

Finally, we examine in more detail the contribution of the mass
produced by all bursts and by bursts in the bright SMG phase as a
function of stellar mass.  In Fig.~\ref{FracSMGmst} we plot the
fraction of the present stellar mass (allowing for recycling) produced
by all bursts in all galaxies (red) and only in descendants of bright
SMGs (black). We also plot the contribution to the stellar mass
produced by bursts and quiescent star formation in the bright SMG
phase for all present-day galaxies (green) and only for the
descendants of bright SMGs (blue).  We see from this plot that on
average bursts contribute 3-8\% to the present-day stellar masses of
galaxies over the whole range $\sim 10^9-10^{12}\hMsol$. If we look
only at galaxies which are descendants of bright SMGs, then we find an
average fraction that is the same at the highest stellar masses, but
increases with decreasing mass, reaching 60\% for the least massive
descendants ($\sim 5\times 10^9 \hMsol$). However, the total stellar
mass produced in the bright SMG phase at $z>1$ is less than this. For
descendants of bright SMGs, the fraction varies from 30-40\% for the
least massive descendants ($\sim 5\times 10^9 \hMsol$), to 0.2-0.4\%
for the most massive ones ($\sim 10^{12}\hMsol$). The average
contribution to the current stellar mass of all galaxies from bursts
in the bright SMG phase increases with stellar mass, but is always
below 0.5\%. For a present-day stellar mass of $1.5\times
10^{11}\hMsol$, which is the median descendant mass of bright SMGs
(see Sec.~\ref{ssec:mass_desc}), the contribution of stellar mass
produced in the bright SMG phase to the present stellar mass is only
1-2\% even for descendants of bright SMGs, and falls to only 0.3\% if
we look at all galaxies of that mass.

\section{Conclusions}

We have investigated predictions for the properties and descendants of
submillimetre galaxies (SMGs) in a theoretical model of galaxy
formation in a $\Lambda$CDM universe \citep{Baugh05}. This model,
which incorporates a top-heavy IMF in starbursts, has previously been
shown to reproduce the observed number counts and redshifts of SMGs,
as well as a wide variety of other galaxy properties at low and high
redshift. We emphasize that \citeauthor{Baugh05} found that the
inclusion of a top-heavy IMF in starbursts was essential in order to
match both the numbers and redshifts of SMGs and the present-day
galaxy luminosity functions in a single model. If instead the IMF was
assumed to be universal and of solar neighbourhood form, then by
adjusting parameters for feedback and other processes, either the
present-day galaxy luminosity functions could be reproduced, but the
sub-mm counts underpredicted, or the sub-mm counts could be
reproduced, but at the price of predicting far too many high
luminosity (and high stellar mass) galaxies at the present day.

We find that bright SMGs in the model, which we define to be objects
having 850$\mum$ fluxes brighter than 5~mJy at some redshift $z>1$,
are typically the progenitors of massive galaxies today, with a median
stellar mass $\sim 2\times 10^{11}h^{-1}M_{\odot}$ {for the
  descendants}. The present-day descendants of these bright SMGs are
predicted to be mainly bulge-dominated systems, but the bright SMG
phase does not appear to play a dominant role in building the bulge,
since their morphologies are similar to those of other present-day
galaxies with similar stellar masses. The descendants of bright SMGs
are also predicted to live in the most massive haloes today. Only the
most massive objects in the present-day universe are predicted to be
predominantly descendants of bright SMGs -- we find that more than
50\% of galaxies with masses above $6 \times 10^{11}h^{-1}M_{\odot}$
and halos with masses above $10^{14}h^{-1}M_{\odot}$ have one of more
progenitors which hosted bright SMGs at some stage in their evolution.

{The model also predicts that the typical stellar masses of
  bright SMGs at $z=2$ are $\sim 2\times 10^{10}h^{-1}M_{\odot}$,
  similar to the new observational estimates by \citet{Hainline10},
  and that they are typically mildly bulge-dominated in stellar mass,
  consistent with the observations by \citet{Swinbank10}. The typical
  duration of the bright SMG phase is predicted to be $\sim 0.1 \Gyr$,
  again consistent with observational estimates.}

In our model, starbursts account for around 30\% of all star formation
over the history of the universe. However, stars formed in galaxies
which are detectable as bright SMGs at $z>1$ are predicted to account
for only 0.06\% of the stellar mass density (in live stars and stellar
remnants) today. There are two reasons why the latter fraction is so
small. Firstly, the recycled fraction for the top-heavy IMF (i.e. the
fraction of the initial stellar mass which is returned as gas to the
ISM when stars die) is $>90\%$, much larger than for a solar
neighbourhood IMF, so that less than 10\% of the stellar mass formed
in bursts remains in stars and remnants today. Secondly, only a small
fraction of ongoing starbursts at $z>1$ have high enough star
formation rates and luminosities to be detectable as bright SMGs. For
present-day galaxies which are descended from bright SMGs at $z>1$,
the fraction of the current stellar mass which was formed in the
bright SMG phase is larger, ranging from $\sim 30\%$ in low mass
descendants to $\sim 0.3\%$ in the highest mass descendants. However,
for median mass descendants, the mean fraction formed in the bright
SMG phase is still only $\sim 3\%$.

In summary, our model predicts that SMGs are the progenitors of
massive galaxies today. However, most of the stellar mass in these
systems is built up by {quiescent star formation and then
  assembled in galaxy mergers}, making the contribution of long-lived
stars formed during the SMG phase typically very small. {The
  latter is contrary to the view commonly expressed in the
  observational literature on SMGs. Thus in our model, the SMG phase
  mainly serves as a ``beacon'' for gas-rich mergers in the
  progenitors of massive galaxies at high-z.}

\section*{Acknowledgements} 
{We thank the referee for their constructive report and
  suggestions which helped to improve the paper.} JEG acknowledges
  receipt of a fellowship funded by the European Commission's
  Framework Programme 6, through the Marie Curie Early Stage Training
  project MEST-CT-2005-021074. This work was supported in part by the
  Science and Technology Facilities Council rolling grant to the
  ICC. CSF acknowledges a Royal Society Wolfson Research Grant Award.
  \bibliographystyle{mn2e} \bibliography{refs}

\begin{thebibliography}{44}
\expandafter\ifx\csname natexlab\endcsname\relax\def\natexlab#1{#1}\fi

\bibitem[{{Almeida} {et~al.}(2010){Almeida}, {Baugh}, \& {Lacey}}]{Almeida10b}
{Almeida} C., {Baugh} C.~M., {Lacey} C.~G., 2010, ArXiv e-prints
  (ArXiv:1011.2300)

\bibitem[{{Baugh}(2006)}]{Baugh06}
{Baugh} C.~M., 2006, Reports on Progress in Physics, 69, 3101

\bibitem[{{Baugh} {et~al.}(2005){Baugh}, {Lacey}, {Frenk}, {Granato}, {Silva},
  {Bressan}, {Benson}, \& {Cole}}]{Baugh05}
{Baugh} C.~M., {Lacey} C.~G., {Frenk} C.~S., {Granato} G.~L., {Silva} L.,
  {Bressan} A., {Benson} A.~J., {Cole} S., 2005, \mnras, 356, 1191

\bibitem[{{Benson} \& {Bower}(2010)}]{Benson10}
{Benson} A.~J., {Bower} R., 2010, ArXiv e-prints

\bibitem[{{Benson} {et~al.}(2003){Benson}, {Bower}, {Frenk}, {Lacey}, {Baugh},
  \& {Cole}}]{Benson03}
{Benson} A.~J., {Bower} R.~G., {Frenk} C.~S., {Lacey} C.~G., {Baugh} C.~M.,
  {Cole} S., 2003, \apj, 599, 38

\bibitem[{{Birnboim} \& {Dekel}(2003)}]{Birnboim03}
{Birnboim} Y., {Dekel} A., 2003, \mnras, 345, 349

\bibitem[{{Bower} {et~al.}(2006){Bower}, {Benson}, {Malbon}, {Helly}, {Frenk},
  {Baugh}, {Cole}, \& {Lacey}}]{Bower06}
{Bower} R.~G., {Benson} A.~J., {Malbon} R., {Helly} J.~C., {Frenk} C.~S.,
  {Baugh} C.~M., {Cole} S., {Lacey} C.~G., 2006, \mnras, 370, 645

\bibitem[{{Chapman} {et~al.}(2003){Chapman}, {Blain}, {Ivison}, \&
  {Smail}}]{Chapman03}
{Chapman} S.~C., {Blain} A.~W., {Ivison} R.~J., {Smail} I.~R., 2003, \nat, 422,
  695

\bibitem[{{Chapman} {et~al.}(2005){Chapman}, {Blain}, {Smail}, \&
  {Ivison}}]{Chapman05}
{Chapman} S.~C., {Blain} A.~W., {Smail} I., {Ivison} R.~J., 2005, \apj, 622,
  772

\bibitem[{{Chapman} {et~al.}(2004){Chapman}, {Smail}, {Windhorst}, {Muxlow}, \&
  {Ivison}}]{Chapman04}
{Chapman} S.~C., {Smail} I., {Windhorst} R., {Muxlow} T., {Ivison} R.~J., 2004,
  \apj, 611, 732

\bibitem[{{Cole} {et~al.}(2000){Cole}, {Lacey}, {Baugh}, \& {Frenk}}]{Cole00}
{Cole} S., {Lacey} C.~G., {Baugh} C.~M., {Frenk} C.~S., 2000, \mnras, 319, 168

\bibitem[{{Coppin} {et~al.}(2008){Coppin}, {Halpern}, {Scott}, {Borys},
  {Dunlop}, {Dunne}, {Ivison}, \& {Wagget al.}}]{Coppin08}
{Coppin} K., {Halpern} M., {Scott} D., {Borys} C., {Dunlop} J., {Dunne} L.,
  {Ivison} R., {Wagget al.}, 2008, \mnras, 384, 1597

\bibitem[{{Dav{\'e}} {et~al.}(2010){Dav{\'e}}, {Finlator}, {Oppenheimer},
  {Fardal}, {Katz}, {Kere{\v s}}, \& {Weinberg}}]{Dave10}
{Dav{\'e}} R., {Finlator} K., {Oppenheimer} B.~D., {Fardal} M., {Katz} N.,
  {Kere{\v s}} D., {Weinberg} D.~H., 2010, \mnras, 404, 1355

\bibitem[{{Dekel} {et~al.}(2009){Dekel}, {Birnboim}, {Engel}, {Freundlich},
  {Goerdt}, {Mumcuoglu}, {Neistein}, {Pichon}, {Teyssier}, \&
  {Zinger}}]{Dekel09}
{Dekel} A., {Birnboim} Y., {Engel} G., {Freundlich} J., {Goerdt} T.,
  {Mumcuoglu} M., {Neistein} E., {Pichon} C., {Teyssier} R., {Zinger} E., 2009,
  \nat, 457, 451

\bibitem[{{Fardal} {et~al.}(2007){Fardal}, {Katz}, {Weinberg}, \&
  {Dav{\'e}}}]{Fardal07}
{Fardal} M.~A., {Katz} N., {Weinberg} D.~H., {Dav{\'e}} R., 2007, \mnras, 379,
  985

\bibitem[{{Font} {et~al.}(2008){Font}, {Bower}, {McCarthy}, {Benson}, {Frenk},
  {Helly}, {Lacey}, {Baugh}, \& {Cole}}]{Font08}
{Font} A.~S., {Bower} R.~G., {McCarthy} I.~G., {Benson} A.~J., {Frenk} C.~S.,
  {Helly} J.~C., {Lacey} C.~G., {Baugh} C.~M., {Cole} S., 2008, \mnras, 389,
  1619

\bibitem[{{Fontanot} {et~al.}(2007){Fontanot}, {Monaco}, {Silva}, \&
  {Grazian}}]{Fontanot07}
{Fontanot} F., {Monaco} P., {Silva} L., {Grazian} A., 2007, \mnras, 382, 903

\bibitem[{{Gonz{\'a}lez} {et~al.}(2010){Gonz{\'a}lez}, {Lacey}, {Baugh}, \&
  {Frenk}}]{Gonzalez10}
{Gonz{\'a}lez} J.~E., {Lacey} C.~G., {Baugh} C.~M., {Frenk} C.~S., 2010. In
  preparation

\bibitem[{{Gonz{\'a}lez} {et~al.}(2009){Gonz{\'a}lez}, {Lacey}, {Baugh},
  {Frenk}, \& {Benson}}]{Gonzalez09}
{Gonz{\'a}lez} J.~E., {Lacey} C.~G., {Baugh} C.~M., {Frenk} C.~S., {Benson}
  A.~J., 2009, \mnras, 397, 1254

\bibitem[{{Granato} {et~al.}(2000){Granato}, {Lacey}, {Silva}, {Bressan},
  {Baugh}, {Cole}, \& {Frenk}}]{Granato00}
{Granato} G.~L., {Lacey} C.~G., {Silva} L., {Bressan} A., {Baugh} C.~M., {Cole}
  S., {Frenk} C.~S., 2000, \apj, 542, 710

\bibitem[{{Hainline} {et~al.}(2010){Hainline}, {Blain}, {Smail}, {Alexander},
  {Armus}, {Chapman}, \& {Ivison}}]{Hainline10}
{Hainline} L.~J., {Blain} A.~W., {Smail} I., {Alexander} D.~M., {Armus} L.,
  {Chapman} S.~C., {Ivison} R.~J., 2010, ArXiv e-prints

\bibitem[{{Hughes} {et~al.}(1998){Hughes}, {Serjeant}, {Dunlop},
  {Rowan-Robinson}, {Blain}, {Mann}, {Ivison}, {Peacock}, {Efstathiou}, {Gear},
  {Oliver}, {Lawrence}, {Longair}, {Goldschmidt}, \& {Jenness}}]{Hughes98}
{Hughes} D.~H., {Serjeant} S., {Dunlop} J., {Rowan-Robinson} M., {Blain} A.,
  {Mann} R.~G., {Ivison} R., {Peacock} J., {Efstathiou} A., {Gear} W., {Oliver}
  S., {Lawrence} A., {Longair} M., {Goldschmidt} P., {Jenness} T., 1998, \nat,
  394, 241

\bibitem[{{Kaviani} {et~al.}(2003){Kaviani}, {Haehnelt}, \&
  {Kauffmann}}]{Kaviani03}
{Kaviani} A., {Haehnelt} M.~G., {Kauffmann} G., 2003, \mnras, 340, 739

\bibitem[{{Kennicutt}(1983)}]{Kennicutt83}
{Kennicutt} Jr. R.~C., 1983, \apj, 272, 54

\bibitem[{{Kroupa}(2001)}]{Kroupa01}
{Kroupa} P., 2001, \mnras, 322, 231

\bibitem[{{Lacey} \& {Cole}(1993)}]{LaceyCole93}
{Lacey} C., {Cole} S., 1993, \mnras, 262, 627

\bibitem[{{Lacey} {et~al.}(2010{\natexlab{a}}){Lacey}, {Baugh}, {Frenk}, \&
  {Almeida}}]{Lacey10b}
{Lacey} C.~G., {Baugh} C.~M., {Frenk} C.~S., {Almeida} C., 2010{\natexlab{a}}.
  In preparation

\bibitem[{{Lacey} {et~al.}(2010{\natexlab{b}}){Lacey}, {Baugh}, {Frenk}, \&
  {Benson}}]{Lacey10a}
{Lacey} C.~G., {Baugh} C.~M., {Frenk} C.~S., {Benson} A.~J.,
  2010{\natexlab{b}}, ArXiv e-prints (ArXiv:1004.3545)

\bibitem[{{Lacey} {et~al.}(2010{\natexlab{c}}){Lacey}, {Baugh}, {Frenk},
  {Benson}, {Orsi}, {Silva}, {Granato}, \& {Bressan}}]{Lacey10}
{Lacey} C.~G., {Baugh} C.~M., {Frenk} C.~S., {Benson} A.~J., {Orsi} A., {Silva}
  L., {Granato} G.~L., {Bressan} A., 2010{\natexlab{c}}, \mnras, 443

\bibitem[{{Lacey} {et~al.}(2008){Lacey}, {Baugh}, {Frenk}, {Silva}, {Granato},
  \& {Bressan}}]{Lacey08}
{Lacey} C.~G., {Baugh} C.~M., {Frenk} C.~S., {Silva} L., {Granato} G.~L.,
  {Bressan} A., 2008, \mnras, 385, 1155

\bibitem[{{Micha{\l}owski} {et~al.}(2010){Micha{\l}owski}, {Hjorth}, \&
  {Watson}}]{Michalowski10}
{Micha{\l}owski} M., {Hjorth} J., {Watson} D., 2010, \aap, 514, A67+

\bibitem[{{Parkinson} {et~al.}(2008){Parkinson}, {Cole}, \&
  {Helly}}]{Parkinson08}
{Parkinson} H., {Cole} S., {Helly} J., 2008, \mnras, 383, 557

\bibitem[{{Press} \& {Schechter}(1974)}]{PressSchechter74}
{Press} W.~H., {Schechter} P., 1974, \apj, 187, 425

\bibitem[{{Puget} {et~al.}(1996){Puget}, {Abergel}, {Bernard}, {Boulanger},
  {Burton}, {Desert}, \& {Hartmann}}]{Puget96}
{Puget} J.-L., {Abergel} A., {Bernard} J.-P., {Boulanger} F., {Burton} W.~B.,
  {Desert} F.-X., {Hartmann} D., 1996, \aap, 308, L5+

\bibitem[{{Schurer} {et~al.}(2009){Schurer}, {Calura}, {Silva}, {Pipino},
  {Granato}, {Matteucci}, \& {Maiolino}}]{Schurer09}
{Schurer} A., {Calura} F., {Silva} L., {Pipino} A., {Granato} G.~L.,
  {Matteucci} F., {Maiolino} R., 2009, \mnras, 394, 2001

\bibitem[{{Silva} {et~al.}(1998){Silva}, {Granato}, {Bressan}, \&
  {Danese}}]{Silva98}
{Silva} L., {Granato} G.~L., {Bressan} A., {Danese} L., 1998, \apj, 509, 103

\bibitem[{{Smail} {et~al.}(2003){Smail}, {Chapman}, {Ivison}, {Blain},
  {Takata}, {Heckman}, {Dunlop}, \& {Sekiguchi}}]{Smail03}
{Smail} I., {Chapman} S.~C., {Ivison} R.~J., {Blain} A.~W., {Takata} T.,
  {Heckman} T.~M., {Dunlop} J.~S., {Sekiguchi} K., 2003, \mnras, 342, 1185

\bibitem[{{Smail} {et~al.}(1997){Smail}, {Ivison}, \& {Blain}}]{Smail97}
{Smail} I., {Ivison} R.~J., {Blain} A.~W., 1997, \apjl, 490, L5+

\bibitem[{{Springel} {et~al.}(2005){Springel}, {White}, {Jenkins}, {Frenk},
  {Yoshida}, {Gao}, {Navarro}, \& {Thacker et al.}}]{Springel05}
{Springel} V., {White} S.~D.~M., {Jenkins} A., {Frenk} C.~S., {Yoshida} N.,
  {Gao} L., {Navarro} J., {Thacker et al.}, 2005, \nat, 435, 629

\bibitem[{{Swinbank} {et~al.}(2008){Swinbank}, {Lacey}, {Smail}, {Baugh},
  {Frenk}, {Blain}, {Chapman}, {Coppin}, {Ivison}, {Gonzalez}, \&
  {Hainline}}]{Swinbank08}
{Swinbank} A.~M., {Lacey} C.~G., {Smail} I., {Baugh} C.~M., {Frenk} C.~S.,
  {Blain} A.~W., {Chapman} S.~C., {Coppin} K.~E.~K., {Ivison} R.~J., {Gonzalez}
  J.~E., {Hainline} L.~J., 2008, \mnras, 391, 420

\bibitem[{{Swinbank} {et~al.}(2010){Swinbank}, {Smail}, {Chapman}, {Borys},
  {Alexander}, {Blain}, {Conselice}, {Hainline}, \& {Ivison}}]{Swinbank10}
{Swinbank} A.~M., {Smail} I., {Chapman} S.~C., {Borys} C., {Alexander} D.~M.,
  {Blain} A.~W., {Conselice} C.~J., {Hainline} L.~J., {Ivison} R.~J., 2010,
  \mnras, 405, 234

\bibitem[{{Tacconi} {et~al.}(2006){Tacconi}, {Neri}, {Chapman}, {Genzel},
  {Smail}, {Ivison}, {Bertoldi}, {Blain}, {Cox}, {Greve}, \&
  {Omont}}]{Tacconi06}
{Tacconi} L.~J., {Neri} R., {Chapman} S.~C., {Genzel} R., {Smail} I., {Ivison}
  R.~J., {Bertoldi} F., {Blain} A., {Cox} P., {Greve} T., {Omont} A., 2006,
  \apj, 640, 228

\bibitem[{{van Dokkum}(2008)}]{vanDokkum08}
{van Dokkum} P.~G., 2008, \apj, 674, 29

\bibitem[{{Vega} {et~al.}(2008){Vega}, {Clemens}, {Bressan}, {Granato},
  {Silva}, \& {Panuzzo}}]{Vega08}
{Vega} O., {Clemens} M.~S., {Bressan} A., {Granato} G.~L., {Silva} L.,
  {Panuzzo} P., 2008, \aap, 484, 631

\end{thebibliography}

\bsp

\label{lastpage}

\end{document}